\documentclass[12pt,american]{scrartcl}
\usepackage{lmodern}

\usepackage{babel}
\usepackage{soul}
\usepackage{xcolor}
\usepackage{geometry}
\usepackage{listings}
\usepackage{color}
\usepackage{verbatim}
\usepackage{amsmath}
\usepackage{setspace}
\usepackage{caption}
\usepackage{subcaption}
\usepackage{esint}
\usepackage{amsfonts}
\usepackage{amssymb}
\usepackage{hhline}
\usepackage[colorlinks=true,citecolor=blue,linkcolor=black,urlcolor=gray]{hyperref}
\usepackage{dcolumn}
\usepackage{float}
\usepackage{wasysym}
\usepackage{booktabs}
\usepackage{graphicx}

\usepackage[normalem]{ulem}
\usepackage{natbib}
\usepackage{cleveref}

\usepackage[linesnumbered,ruled]{algorithm2e}
\makeatletter
\newcommand{\algorithmfootnote}[2][\footnotesize]{%
	\let\old@algocf@finish\@algocf@finish
	\def\@algocf@finish{\old@algocf@finish
		\leavevmode\rlap{\begin{minipage}{\linewidth}
				#1#2
		\end{minipage}}%
	}%
}
\makeatother

\SetAlgoLined

\definecolor{dkgreen}{rgb}{0,0.6,0}
\definecolor{gray}{rgb}{0.5,0.5,0.5}
\definecolor{mauve}{rgb}{0.58,0,0.82}

\makeatletter
\usepackage[utf8]{inputenc}

\title{OPUS: An Integrated Assessment Model for Satellites and Orbital Debris\thanks{This work was supported by 2022 grant funding from the NASA ROSES program.}}

\author{
    Akhil Rao\textsuperscript{1} \and
    Mark Moretto\textsuperscript{2} \and
    Marcus Holzinger\textsuperscript{2} \and
    Daniel Kaffine\textsuperscript{3} \and
    Brian Weeden\textsuperscript{4}
}

\date{\today}

\begin{document}

\newgeometry{verbose,tmargin=1cm,bmargin=1.5cm,lmargin=1in,rmargin=1in,headheight=1cm,headsep=1cm,footskip=0.5cm}

\maketitle

\setcounter{footnote}{0}
\renewcommand{\thefootnote}{\arabic{footnote}}

\footnotetext[1]{Middlebury College, Department of Economics}
\footnotetext[2]{University of Colorado Boulder, Ann and H. J. Smead Aerospace Engineering Sciences}
\footnotetext[3]{University of Colorado Boulder, Department of Economics}
\footnotetext[4]{Secure World Foundation}

\setcounter{footnote}{0}

\begin{abstract}
     An increasingly salient public policy challenge is how to manage the growing number of satellites in orbit, including large constellations. Many policy initiatives have been proposed that attempt to address the problem from different angles, but there is a paucity of analytical tools to help policymakers evaluate the efficacy of these different proposals and any potential counterproductive outcomes. To help address this problem, this paper summarizes work done to develop an experimental integrated assessment model---Orbital Debris Propagators Unified with Economic Systems (OPUS)---that combines both astrodynamics of the orbital population and economic behavior of space actors. For a given set of parameters, the model first utilizes a given astrodynamic propagator to assess the state of objects in orbit. It then uses a set of user-defined economic and policy parameters---e.g. launch prices, disposal regulations---to model how actors will respond to the economic incentives created by a given scenario. For the purposes of testing, the MIT Orbital Capacity Tool (MOCAT) version 4S was used as the primary astrodynamics propagator to simulate the true expected target collision probability ($p_c$) for a given end-of-life (EOL) disposal plan. 
     To demonstrate propagator-agnosticism, a Gaussian mixture probability hypothesis density (GMPHD) filter was also used to  simulate $p_c$.
     We also explore economic policy instruments to improve both sustainability of and economic welfare from orbit use. In doing so, we demonstrate that this hybrid approach can serve as a useful tool for evaluating policy proposals for managing orbital congestion. We also discuss areas where this work can be made more robust and expanded to include additional policy considerations.
\end{abstract}

\restoregeometry

\section{Background}

Over the last ten years, there has been rapid growth in the number of satellites launched into Earth orbit. Today, more than 8,700 active satellites in Earth orbit, along with tens of thousands of pieces of debris left over from the several previous decades of space activities. Commercial companies and governments have announced plans to place tens to hundreds of thousands of additional satellites in orbit over the next decade, raising increasing concerns about the ability of policymakers to effectively manage traffic in orbit and reduce the probability of catastrophic collisions between satellites and orbital debris.

The growing awareness of this situations has sparked debate over a wide variety of policy proposals for mitigating this risk. These proposals include orbital slotting concepts that would better organize where satellites can be placed, tightening the existing international standard of 25 years for post-mission disposal (PMD), mandatory propulsion on all future satellites, actively removing existing large debris objects, and various types of orbital use fees. While many of these proposals have merits on paper, policymakers have few analytical tools to evaluate their efficacy. More importantly, the existing tools tend to focus only on the astrodynamics of how objects might interact in orbit and do not include potential changes to the behaviors of the entities controlling current and planning future space activities, particularly their potential response to economic incentives. 

To address this situation, we proposed development of an experimental hybrid model that incorporates both physics and economic models. This Integrated Assessment Model (IAM) would be able to combine both the astrodynamic behavior of space objects on orbit and the economic behavior of their controlling entities on Earth, thus giving policymakers the ability to more robustly assess various public policy proposals. While still in its early stages, we believe OPUS---Orbital Debris Propagators Unified with Economic Systems---is a useful approach that will provide significant assistance to policymakers in deciding how best to mitigate the space sustainability challenges stemming from the growing number of satellites and large constellations in orbit. Unlike prior work on orbital-use IAMs using atheoretical/purely-empirical economic models of launch behavior \citep{raoletizia2021}, the IAM developed here uses a theoretically-grounded economic model of launch behavior. Grounding launch behavior in economic theory enables a richer variety of counterfactual and policy analyses.

The following sections describe the details of both the physics and economics models that make up the hybrid model we developed, along with some initial results to demonstrate the outputs from the model. The results presented here do not rely on a specific propagator, parameterization of orbital locations, or classification of orbital object types. Where necessary, MOCAT-4S location parameterization (``altitude shells'') and object classifications (``slotted, unslotted, debris, and derelicts'') are used. ``Satellites'' is used as a generic term for returns-producing orbiting objects, and ``debris'' is used as a generic term for all other orbiting objects. Section \ref{sec:model-architecture} provides an overview of the OPUS framework, with Sections \ref{sec:debris-environment-model} and \ref{sec:econ-behavior} describing the debris environment and economic behavior models in more detail. The economic behavior model is the focus. Section \ref{sec:results} presents results from several scenarios to illustrate OPUS' capabilities, and Section \ref{sec:future} provides some directions for future work. We conclude in section \ref{sec:conclusion}.

\section{Model architecture}
\label{sec:model-architecture}

OPUS involves two coupled models interacting to determine at each time step (a) the state of the orbital environment, and (b) launch rates to different locations in the orbital environment. This architecture is tailored to achieve three outcomes:
\begin{enumerate}
    \item to provide insight into economic and technical factors that may exacerbate or yield potential solutions to the debris problem;
    \item to remain agnostic to the specific propagator used so as to maximize compatibility with existing analytical tools and workflows for debris environment analysis;
    \item to enable sensitivity analysis over key parameters and proposed policy designs.
\end{enumerate}

This prototype is written in the MATLAB and R programming languages. Model simulation and propagation conducted in MATLAB and analytics and figures of merit are generated in R.

OPUS is written for situations where a constellation is operating in a particular orbital volume with other satellite operators (the ``competitive fringe''). Operators in the fringe each control relatively few satellites compared to the constellation operator and are assumed to behave according to a system of ``open-access conditions''. The open-access conditions are the key innovation in OPUS.

Economic models of behavior use constrained optimization problems to reflect interactions between purposeful agents attempting to achieve their objectives subject to various limitations. When property rights over a desirable resource are absent but one agent’s use detracts from another agent’s use, the solution to the multi-agent constrained optimization problem can be reduced to an ``open-access condition'' which reflects the fact that agents will only stop changing their behaviors once there are no further economic profits to doing so \citep{gordon1954economic,libecapwiggins1984}. The open-access condition defines feedback rules linking the state of the resource and operating environment to agents’ choices. Existing economic literature on the orbital debris problem notes that Article II of the Outer Space Treaty, combined with the possibility of physical congestion, imply that commercial use of orbital space will be governed by a system of open-access conditions \citep{Adilov2015,rao2020orbital,rouillon2020physico,rao2023economics}.

\begin{figure}[htbp]
    \centering
    \includegraphics[width=\textwidth]{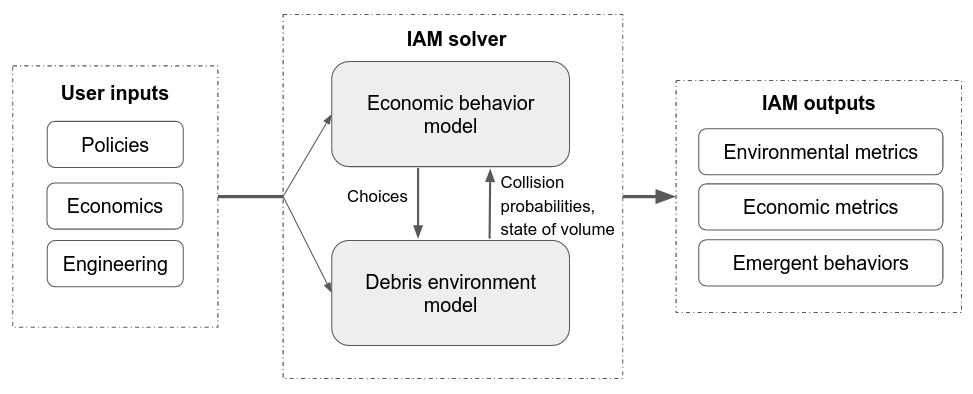}
    \caption{Schematic diagram outlining OPUS.}
    \label{fig:iam-schematic}
\end{figure}

The constellation operators' launch plans are set exogenously by the user, along with parameters applying to both the constellation operators and the fringe such as EOL disposal requirements and compliance rates. By default OPUS assumes there are two constellations: a larger system near 550 km altitude and a smaller system near 1100 km altitude. The constellations are described in more detail in Section \ref{sec:econ-behavior}. Both the constellation and fringe operators' behaviors alter the state of the debris environment and incent behavioral responses from the competitive fringe. The economic model uses the open-access condition to project these behavioral responses from the fringe. Figure \ref{fig:iam-schematic} illustrates the inputs and interactions in the model in a high-level schematic diagram.

The user determines updates to key economic, physical, and policy parameters for the desired scenario in CSV files. The model begins from the initial population files used for MOCAT-4S simulations, reflecting the state of the orbital environment circa July 2022. A bash script orchestrates the desired series of scenarios and allows the user to set the desired simulation horizon. By default the script compares two types of launch behavior: ``satellite feedback'' behavior which maintains the initial population, and ``equilibrium'' behavior which uses the open-access conditions to determine launch patterns. OPUS operates in one-year time steps.

The minimal mode of operation involves the user retaining default parameters except for desired changes listed in a CSV file. Default parameter values and examples of parameter changes are shown in Section \ref{sec:model-parameter-tables}. The OPUS repository provides examples of CSVs for scenarios shown in Section \ref{sec:results}. 

\subsection{Debris environment model}
\label{sec:debris-environment-model}

Consider a set of $K$ orbital locations (``orbits''), $k = 1, \ldots, K$, where each location may be an altitude shell (``orbital''), an altitude-inclination bin, or some other parameterization of groups of paths in LEO. At each location, there are $S_{it}$ satellites of type $i \in I$ and $D_{jt}$ debris objects of type $j \in J$ in period $t$. Let $\cdot$ subscripts represent arrays over the subscripted index, e.g. $S_{\cdot t} = [S_{it}]_i$ is a vector of stocks of all satellite types at time $t$. 

The number of active satellites of type $i$ in location $k$ and orbit in period $t+1$ ($S_{i k t+1}$) is the number of launches of type $i$ in location $k$ in the previous period ($X_{itk}$) plus the number of satellites which survived the previous period ($\mathcal{S}_{ik}(S_{\cdot \cdot t}, D_{\cdot \cdot t})$, where $\mathcal{S}_{ik}$ are physical dynamics mapping satellite stocks of type $i$ at location $k$ and $\mathcal{S} = [\mathcal{S}_{ik}]_i$ is the associated vector of next-period satellite stocks). The amount of debris of type $j$ in location $k$ and orbit in $t+1$ ($D_{j k t+1}$) is the net debris remaining after orbital decay and fragment formation processes ($\mathcal{D}_{jk}(S_{\cdot \cdot t},D_{\cdot \cdot t})$), plus the amount of debris in the shell created by new launches ($\sum_i m_{ijk} X_{itk}$). The laws of motion for the satellite and debris stocks in each location are:
\begin{align}
    \label{eqn:sat-lom}
    S_{i k t+1} &= \mathcal{S}_{ik}(S_{\cdot \cdot t}, D_{\cdot \cdot t}) + X_{itk} \\
    \label{eqn:deb-lom}
    D_{j k t+1} &= \mathcal{D}_{jk}(S_{\cdot \cdot t}, D_{\cdot \cdot t}) + \sum_{i} m_{ijk} X_{itk} .
\end{align}

Equations \ref{eqn:sat-lom} and \ref{eqn:deb-lom} define general laws of motion for propagating the state of the orbital environment, and may be implemented by a variety of different propagators.

\subsubsection{Propagators}
\label{sec:propagator-descriptions}

OPUS currently implements the physical dynamics functions $\mathcal{S}_{ik}$ and $\mathcal{D}_{jk}$ using MOCAT-4S or a GMPHD filter as propagators. We describe each below.

\paragraph{MOCAT-4S.} 
The primary propagator used in OPUS is MOCAT-4S, a source-sink evolutionary model (SSEM) of low-Earth orbit. MOCAT itself describes a family of SSEMs used to assess orbital capacity under different object configurations. The 4S version, used here, models the evolution of four types of objects---slotted active satellites, unslotted active satellites, derelict intact objects, and small fragments---in a set of ordinary differential equations. Active satellites and intact derelicts are assumed to be physically identical. MOCAT-4S discretizes the region between 200-1600 km above mean sea level into 40 non-overlapping shells of 35 km each, so that the 40 shells are the $K$ locations used in OPUS. It uses the NASA Standard Breakup Model to simulate catastrophic and non-catastrophic explosions, and propagates the state of the environment at annual timesteps. 
MOCAT-4S is open-source and computationally tractable, with a single year's propagation taking on the order of seconds.

The slotted and unslotted satellite populations correspond to the active satellites shown in equation \ref{eqn:sat-lom}, while the derelicts and fragments correspond to the debris objects shown in equation \ref{eqn:deb-lom}. `Slotting'', in the terminology of the MOCAT family of models, refers to whether a satellite's orbital parameters have been configured to minimize collisions with other slotted satellites \citep{arnas2021definition}. MOCAT has so far been primarily used to assess orbital capacity to sustainably hold satellites under different degrees of slotting effectiveness and adoption \citep{d2022capacity, lifson2022many, jang2022stability}. While orbital capacity in this sense it is not the focus here, the distinction between slotted and unslotted is convenient for our purposes. Slotted satellites are mapped to constellations operated by a single entity, while unslotted satellites are mapped to the open-access fringe. These distinctions are described in more detail in Section \ref{sec:econ-behavior}.

\paragraph{GMPHD filter.}
To demonstrate OPUS' propagator-agnosticism, a GMPHD filter is used to propagate the state of the LEO environment as well. GMPHD filters are generic statistical tools to propagate high-dimensional states under tracking uncertainty via mixtures of Gaussian distributions. Each Gaussian in the mixture represents a particular hypothesis about the state of the system being modeled, e.g. the location and size of a group of debris objects. These hypotheses include the covariance structure between the states being tracked to capture uncertainties and their correlations. A unique feature of GMPHD filters is that the integral of each Gaussian is the number of objects being tracked by that hypothesis (rather than simply integrating to one). As the system evolves, the filter splits and merges Gaussians to represent the degree of uncertainty in the system of hypotheses, subject to user-chosen computational considerations. GMPHD filters have been used in a range of multi-target tracking problems, including astrodynamics and robotics \citep{mashiku2012statistical, huang2022estimations}, though to our knowledge it has not been used to implement a SSEM of the LEO environment.

We parameterize the GMPHD filter with three types of objects: constellation and open-access fringe satellites (equation \ref{eqn:sat-lom}), and the debris population \ref{eqn:deb-lom}. For simplicity, the satellites are identical and assumed to have no self-collisions with any other satellites, i.e. perfect universal slotting. The debris population is parameterized as a continuous family of objects with sizes ranging from 0-1 m. We parameterize altitude as a continuous range, so that debris objects are described by two infinite-dimensional parameters: their location and their size. Satellite objects are launched to discrete bins, similar to MOCAT, with uniform density within the bins. The model is propagated at annual timesteps and collisions are sampled at uniform sub-timesteps within each year. We integrate over altitudes and debris sizes to report summary measures in terms of discrete locations and debris size classes.

\begin{figure}[htbp]
    \centering
    \includegraphics[width=\textwidth]{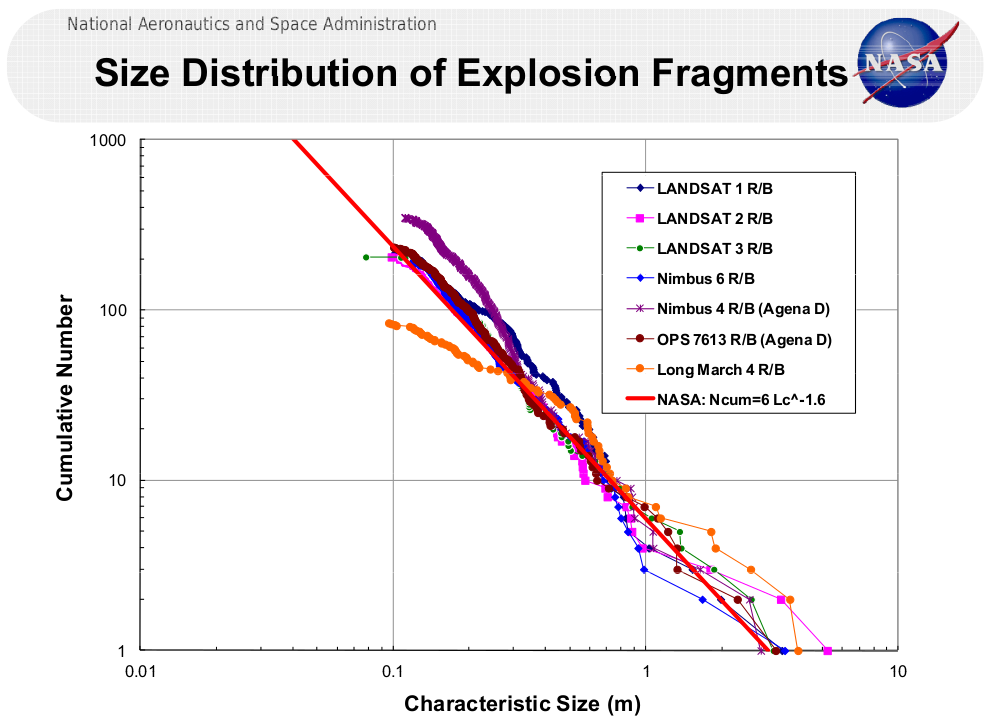}
    \caption{Size distribution of explosion fragments from slide 30/43 of \cite{liou2012orbital}. The prevalence of fragments smaller than the detection limit can be inferred from the parametric fit.}
    \label{fig:lnt-dist-inferred}
\end{figure}

The continuous range of debris over sizes is particularly useful for modeling the evolution of the LEO debris environment, as it enables tracking of lethal non-trackable (LNT) fragments. Though these fragments cannot be directly tracked, their existence can be statistically inferred from the distribution of trackable objects. Figure \ref{fig:lnt-dist-inferred} shows observations of the size distribution of explosion fragments along with a parametric fit predicting fragments below the detection limit \citep{liou2012orbital}. As better understanding of the size distribution of fragments develops, the covariance structure and splitting/merging rates of the GMPHD filter can be reparameterized to better constrain the evolution of LNTs. Runtime of the GMPHD filter can vary substantially depending on computational parameters such as the number of Gaussians used in the mixture, the frequency of splitting/merging, and the number of timesteps at which collisions are sampled. We discuss potential future research directions using infinite-dimensional models such as the GMPHD filter in section \ref{sec:conclusion}.

\subsubsection{EOL disposal}
\label{sec:disposal-implementation}
    
    We implement EOL disposal as immediate movement of satellites at the end of their productive life to the highest altitude consistent with the user-chosen disposal regulation via Hohmann transfer, followed by a recircularization burn. At the start of every simulation, the user may set a desired maximum disposal time regulation $T_D$. This is a time within which an intact satellite must deorbit after its mission is over, i.e. a compliant satellite will be in orbit for mission life plus $T_D$ years. The default disposal time is set to 5 years.
    
    Every disposal time $T_D$ implies a maximum compliant altitude $A_D$. We apply the U.S. Standard Atmosphere and CIRA-72 atmospheric density model, also used in MOCAT-4S with values taken from \cite{david2013vallado}, to compute decay rates and corresponding decay times by altitude.
    
    We assume a fraction $\phi$ of satellites are non-compliant with the regulation and are left where they are at the end of their mission. The default setting is $\phi=0$, i.e. full compliance. If a compliant satellite is at an altitude $a > A_D$, at the end of its mission it is moved to altitude $A_D$ as described above. The transfer is assumed to be instantaneous relative to the scale of model time steps.\footnote{Incorporating elliptical disposal is challenging given that such disposals can significantly increase collision risks across many locations for brief (relative to model timesteps) durations. This is an important step for future work.}

\subsection{Economic behavior model}
\label{sec:econ-behavior}

Consider two types of satellites, constellation satellites ($i=1$) and open-access fringe satellites ($i=2$), serving different industries such as telecommunications and imaging.\footnote{Fringe operators may operate multiple satellites, but at a scale much smaller than the constellation operator so that they can be approximated as owning a single satellite. The constellation index may represent multiple constellations with different operators, though we assume each constellation occupies a single location and does not co-locate with other constellations. These assumptions can be relaxed without any loss of generality.} Define $P_{ik}$ as the probability a satellite of type $i$ collides with another object in location $k$. A satellite in the fringe expects to face a collision probability (from all sources) of $P_{2k}(S_{\cdot \cdot t},D_{\cdot \cdot t})$ across various locations $k$.\footnote{It is convenient to assume $P_{1k}$ is uniformly weakly smaller than $P_{2k}$ since constellation satellites are slotted while fringe satellites are not.} On average, fringe satellites have active lifetimes of $\mu^{-1}$ years.\footnote{If the propagator has separate compartments for active and derelict satellites, the lifetime $\mu^{-1}$ should be consistent between economic and physical models.}  

Under open access, the marginal fringe operator earns zero economic profits at any location they can access.  The net rate of return earned by satellites in the fringe is $R_k(S_{2\cdot t}) - r$, where $r$ is the discount rate representing the opportunity cost of funds, and $R_k(S_{2\cdot t})$ is the gross rate of return earned by the satellite given economic competition within the fringe industry (and any spectrum congestion), and $\tau_{k t}$ is a location-time specific tax rate---``orbital-use fees''---reflecting economic policies imposed to improve sustainability \citep{rao2020orbital}.\footnote{Orbital-use fees in this model setting may be implemented in many ways, e.g. as launch taxes, as satellite deorbiting performance bonds, as direct satellite taxes, etc. All are equivalent here, albeit with differences in implied structure given the same fee rate. See \citep{adilov2023economics} for more on satellite deorbiting performance bonds.} $R_k(S_{2\cdot t})$ is weakly decreasing in $S_{2\cdot t}$. Under open access to orbit in period $t-1$, the fringe will launch $\hat{X}_{2kt-1}$ satellites to the volume until the following system of equations is satisfied across all locations $k$:

\begin{equation}
    \forall k, ~~ \hat{X}_{2kt-1} : R_k(S_{2\cdot t}) - r - \mu - P_{2k}(S_{\cdot \cdot t},D_{\cdot \cdot t}) - \tau_{k t} = 0.
    \label{eqn:open-access-condition-system}
\end{equation}

We assume the constellation's launch plans are publicly announced in advance and are exogenous to the fringe's choices. $P_i$ is calculated from the same model that computes $\mathcal{S}_{i}$ and $\mathcal{D}_k$. The net rate of return function has two components: the expected future revenues or payoffs that the satellite delivers in period $t$, $q_k(S_{2\cdot t})$, and the annualized unit cost of deploying it, $c_k$:
\begin{equation}
    R_k(S_{2 \cdot t}) = \frac{q_k(S_{2\cdot t})}{c_k}.
\end{equation}

To better illustrate how open-access launching operates, we compare it against ``satellite feedback'' behavior, which simply replenishes location-specific populations assuming no collisions.

\paragraph{Revenue function parameterization.} When the location index $k$ indicates altitude bins, we parameterize the revenue function as linear with a common coefficient across all altitudes:
\begin{equation}
    q_k(S_{2 \cdot t}) = \alpha_1^q - \alpha_2^q \sum_{k \in K} S_{2 k t}
\end{equation}

The parameters of the revenue function are set to match the following conditions:
\begin{enumerate}
    \item The maximum willingness-to-pay for service from a fringe satellite is $\alpha_1^q = 7.5 \times 10^{5}$ \$/sat. 
    \item Fringe satellites at all locations are perfect substitutes, and willingness-to-pay for service from a marginal fringe satellite declines at $\alpha_2^q = 100$ \$/sat.
\end{enumerate}

Different rates of substitution across orbits due to different output characteristics---e.g. certain locations being ideal for specific types of imagery collections---could be reflected in $k$-specific coefficients of $q_k(S_{2 k t})$, i.e. $\alpha_2^q \to \alpha^p_{2 k}$. Collecting the right data and estimating that credibly is a task for future research.

\paragraph{Cost function parameterization.} The cost function reflects three factors:
\begin{enumerate}
    \item The lift price, $c_{lift}$. This is the dollar cost per kilogram of accessing LEO, multiplied by the mass of the satellite payload. We set the lift price to \$5,000 per kg following the vehicle-weighted launch price index developed in \cite{corrado2023space}.
    \item The cost of the delta-v budget at a given altitude $k$, $c_{\Delta v}(k)$. This is the total dollar cost of the delta-v (expressed in units of $m/s$) required to maintain the satellite in its target orbit and conduct any necessary maneuvers over its lifetime. Letting $v_{drag}(k)$ be the force exerted on the satellite by atmospheric drag at altitude $k$, $\mu^{-1}$ be the satellite's lifetime in years, $f_s$ be a multiplicative factor reflecting increased drag due to solar flux variations over the satellite's lifetime, and $f_m$ be a safety margin for additional maneuvers in delta-v units, we compute the delta-v budget as:
    \begin{equation}
    f_s \mu^{-1} v_{drag}(k) + f_m.
    \end{equation}
    We set $f_s = 1.5$ and $f_m = 100$. We monetize the delta-v budget at $p_{\Delta v} = \$1,000$ per $m/s$. These values are arbitrarily chosen due to lack of data.
    \item The opportunity cost of lost lifetime due to deorbit from altitude $k$ to altitude $k^\star$, $c_{\mu}(k, k^\star)$. This is the cost of a satellite's lifetime being reduced by expending fuel to deorbit. We keep the satellite's operational lifetime fixed at $\mu^{-1}$ years for simplicity.\footnote{This modeling choice is made for simplicity; future work will extend the model to reflect more realistic lifetime reductions.} We assume the satellite conducts a Hohmann transfer from its initial circular orbit to a circular orbit at the target disposal altitude, followed by a recircularization burn as described in Section \ref{sec:disposal-implementation}. Given the delta-v requirement for this maneuver and the initial delta-v budget, the lifetime reduction is calculated as the lost share of delta-v required for stationkeeping and routine maneuvers, monetized at the maximum willingness-to-pay for satellite service. This provides an upper bound on the opportunity cost of deorbit maneuvers.\footnote{This modeling choice is made for tractability; future work will extend the model to feature dynamically-updating opportunity costs based on changes in annual revenues per satellite, accounting for every other operator's launch and location choice.}
\end{enumerate}

Letting the rate of non-compliance with deorbit regulations be $\phi$, the complete cost function for altitude $k$ given a target deorbit location $k^\star$ is therefore:
\begin{equation}
    c_k = c_{lift} + c_{\Delta v}(k) + (1 - \phi)c_{\mu}(k, k^\star)
\end{equation}

\begin{figure}[htbp]
    \centering
    \begin{subfigure}[b]{0.75\textwidth}
        \centering
        \vspace{-2cm}
        \includegraphics[width=\textwidth]{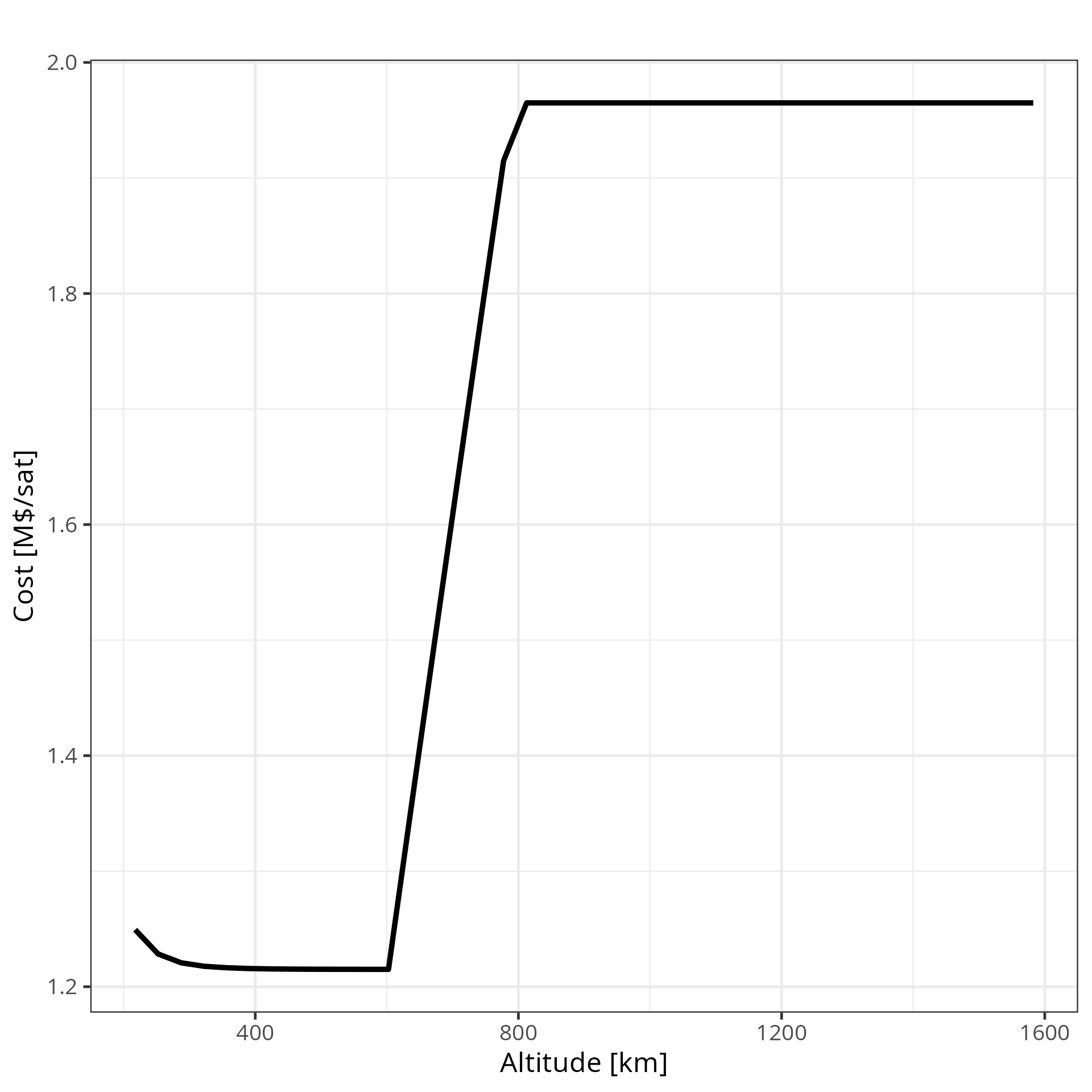}
        \caption{Cost function given 25-year deorbit rule with full compliance.}
        \label{fig:cost-fn-25}
    \end{subfigure}
    
    \vspace{1em}
    
    \begin{subfigure}[b]{0.75\textwidth}
        \centering
        \includegraphics[width=\textwidth]{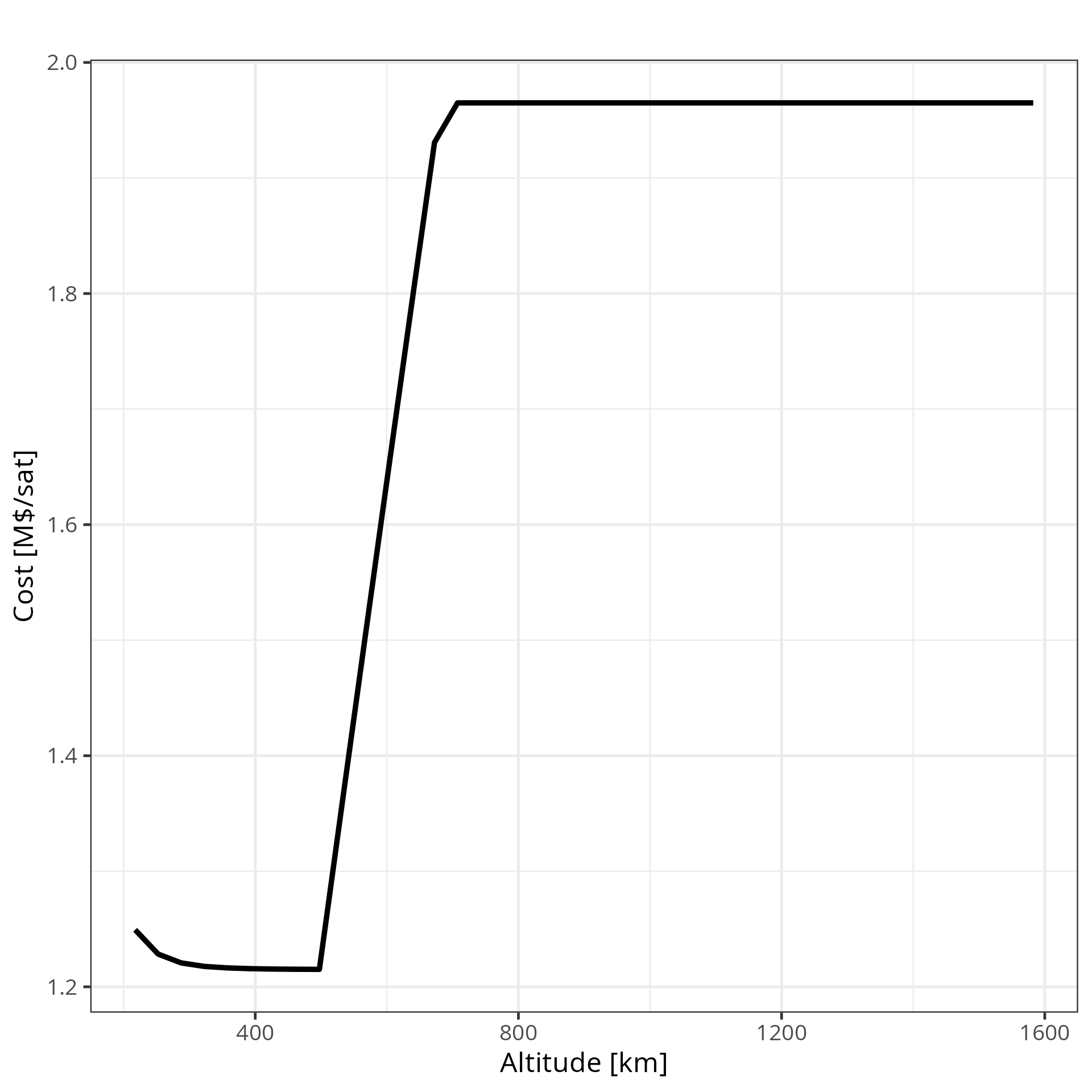}
        \caption{Cost function given 5-year deorbit rule with full compliance.}
        \label{fig:cost-fn}
    \end{subfigure}
    
    \caption{Comparison of cost functions for different deorbit rules.}
    \label{fig:cost-functions}
\end{figure}

The cost functions across altitudes, given 25- and 5-year deorbit times with full compliance, are shown in figure \ref{fig:cost-functions}. The cost functions initially decline over altitude with atmospheric density, reflecting lower stationkeeping costs. Above the maximum 5-year deorbit-compliant altitude, the cost rises steeply, reflecting the cost of the disposal maneuver. The cost flattens once deorbit maneuvers exhaust the satellite's full lifetime delta-v supply.

\subsection{Constellation parameterization} 
\label{sec:constellation-parameterization}

We assume the constellations begin at sizes and locations determined by the initial population files. From there, the user can define the parameters governing the constellation's launch patterns by altering values in \texttt{constellation-parameters.csv} (assumed to be in \texttt{scenarios/parameters/}). The parameters and their defaults are listed below:
\begin{enumerate}
    \item \texttt{n\_constellations}: The number of constellations. Default is 2.
    \item \texttt{location\_indices}: The locations at which constellations are to be built up, expressed as indices under the model parameterization. Defaults are 10 and 29, corresponding to 500 km and 1100 km, i.e. a ``large and low'' constellation and a ``small and high'' constellation. (CHECK)
    \item \texttt{target\_sizes}: The final sizes the constellations seek to reach. The default values are 3000 for the lower constellation and 300 for the higher constellation.
    \item \texttt{max\_launch\_rates}: The maximum launch rates the constellations can achieve, reflecting launch capacity available. Every period the constellation controller will check whether the constellations are at their target sizes and if not, launch up to the maximum launch rate for that constellation to replenish or build up the system. Defaults are 1500 for the lower constellation and 150 for the higher constellation.
\end{enumerate}

\subsection{Model parameters}
\label{sec:model-parameter-tables}

Table \ref{tab:econ-parameters} lists the key economic parameter values in the benchmark scenario, units, with a brief description. We refer readers to the code files for MOCAT-4S and the GMPHD filter for a full list of physical parameters.

\begin{table}[h]
    \centering
    \begin{tabular}{|l|c|c|p{6cm}|}
    \hline
    \textbf{Parameter} & \textbf{Value} & \textbf{Units} & \textbf{Description} \\
    \hline
    $\mu^{-1}$ & 5 & years & Active satellite lifetime \\
    \hline
    $T_D$ & 5 & years & End-of-life disposal regulation \\
    \hline
    $\phi$ & 0 & \%/year & Disposal non-compliance rate \\
    \hline
    $r$ & 5 & \%/year & Discount rate \\
    \hline
    $\alpha^q_1$ & \(7.5 \times 10^5\) & \$/year & Revenue with no competition \\
    \hline
    $\alpha^q_2$ & \(1.0 \times 10^2\) & \$/satellite/year & Linear revenue coefficient \\
    \hline
    $\tau_{kt}$ & 0 & \%/satellite/year & Shell-specific orbital-use fee rate in period $t$ \\
    \hline
    $f_s$ & 1.5 & unitless & Stationkeeping delta-v safety factor \\
    \hline
    $f_m$ & 100 & m/s & Additional delta-v for discretionary maneuvers \\
    \hline
    $p_{\Delta v}$ & 1000 & \$/m/s & Cost of delta-v \\
    \hline
    $c_{lift}$ & 5000 & \$/kg & Launch price per kilogram \\
    \hline
    \end{tabular}
    \caption{Benchmark values of key economic parameters.}
    \label{tab:econ-parameters}
\end{table}

\subsection{Using the model}
\label{sec:using-the-model}

OPUS is divided into a series of MATLAB functions for simulation and R scripts for reporting, with CSV files used to read in user-defined policy, economic, and engineering parameters. The primary mode of operation is via command line interface using a shell (bash) script. The shell script and solver code are described below.

\paragraph{\texttt{conductor.sh}} The shell script \texttt{conductor.sh} calls the various functions and scripts to run simulations. The script is broken into 3 blocks:

\begin{enumerate}
    \item \textbf{Initialization:} Sets up the initial parameters. This is where the user specifies the type of propagator to use, currently either \texttt{MOCAT} or \texttt{GMPHD}.
    \item \textbf{Scenario Setup:} Sets up array of scenarios to be run. The default value is \texttt{benchmark}. CSV files with scenarios should be provided with paths relative to the location of \texttt{conductor.sh}. The default is \texttt{scenarios/parsets/<filename>.csv}. Simulation horizon and number of workers for parallelization are also set here in the \texttt{model\_horizon} and \texttt{n\_workers} variables.
    \item \textbf{Execution and Post-Processing Loop:} The script iteratively executes the MATLAB solver for each scenario, first with \texttt{equilibrium} behavior (launching till zero profits each period) and then with \texttt{sat\_feedback} behavior (launching to maintain populations, ignoring collisions). At the start of each run a unique human-readable name is generated based on the scenario parameters for better traceability and management of the results. \texttt{analytics.R} is run after each propagator-behavior combination to generate figures and CSVs for individual scenarios, and then \texttt{compare-two-scenarios.R} is run after \texttt{analytics.R} to generate images and CSVs comparing pairs of scenarios.
\end{enumerate}

While the main solver can be used independently, \texttt{conductor.sh}, facilitates setting up, executing, and analyzing multiple scenarios in batches without having to individually manage each run.

\paragraph{\texttt{iam\_solver.m}} The main solver file is \texttt{iam\_solver.m}. It takes the following input strings in order to produce CSV files that describe the time evolution of the orbital environment across the MOCAT orbital shell discretization (40 shells of 35 km each between 200 - 1600 km):

\begin{enumerate}
    \item \texttt{stem}: A unique filename for the model run. A folder with this name is created in the \texttt{scenarios} folder, and all model outputs have this prefix. The \texttt{scenarioNamer.m} file generates unique human-readable names based on the model parameters.
    \item \texttt{model\_type}: Choice of propagator. The current implementation allows either \texttt{MOCAT} or \texttt{GMPHD} to be supplied as inputs. Scenario functionality is most thoroughly tested for \texttt{MOCAT}.
    \item \texttt{launch\_pattern\_type}: Choice of launch behavior. The current implementation allows either \texttt{equilibrium} (for open-access equilibrium) or \texttt{sat\_feedback}. By default \texttt{conductor.sh} will run both types of behavior when called. Currently only \texttt{equilibrium} has been tested with both \texttt{MOCAT} and \texttt{GMPHD}.
    \item \texttt{parameter\_file}: Path to CSV file containing any changes to the default parameter values for scenario analysis. The file is expected to be in the \texttt{scenarios/parsets/} folder.
    \item \texttt{model\_horizon\_str}: Number of years to propagate the orbital environment. Can be provided by the user as an int or string in \texttt{conductor.sh}.
    \item \texttt{n\_workers}: Number of workers for solver parallelization when \texttt{equilibrium} launch behavior is selected. Solver code is parallelized across locations, so something close to the number of locations is usually a good choice.
\end{enumerate}

Algorithm \ref{algo:iam-solver} provides a high-level description of the inputs, order of operations, and outputs involved in running OPUS.

\begin{algorithm}
    \caption{Overview of \texttt{iam\_solver.m}}
    \label{algo:iam-solver}
    \KwData{stem, model\_type, launch\_pattern\_type, parameter\_file, model\_horizon\_str, n\_workers}
    \KwResult{Outputs CSV files describing orbital environment evolution}
    
    \textbf{Step 1: Initialization and Parameter Preparation}\;
    \Indp
        Call \texttt{MOCAT4S\_Var\_Cons}() and \texttt{GMPHD\_VAR\_Cons}() to initialize MOCAT and GMPHD constants objects \textbf{VAR} and \textbf{GMPHD\_params}\;
        Call \texttt{set\_econ\_parameters}() with \textbf{VAR} to initialize \textbf{econ\_params}\;
    \Indm
    
    \textbf{Step 2: Modify Parameters for Scenarios}\;
    \Indp
        \If{parameter\_file is provided}{
            Call \texttt{modifyParameters}() for specified scenarios, update \textbf{VAR} and \textbf{econ\_params}\;
        }
    \Indm
    
    \textbf{Step 3: Select and Initialize Propagation Method}\;
    \Indp
        \If{model\_type is MOCAT}{
            Load initial orbital state and set up MOCAT model\;
        }
        \ElseIf{model\_type is GMPHD}{
            Load initial orbital state and set up GMPHD model\;
            Perform one-step propagation to calculate debris distribution over discrete categories: lethal non-trackables, small trackables, large trackables in MOCAT-consistent altitude bins\;
        }
    \Indm
    
    \textbf{Step 4: Build Cost Function}\;
    \Indp
        Call \texttt{buildCostFunction}() to get \textbf{cost\_fn\_params} incorporating disposal regulations (in \textbf{econ\_params}) and compliance rate (in \textbf{VAR})\;
        Update \textbf{econ\_params} and \textbf{VAR} with \textbf{cost\_fn\_params}\;
    \Indm
    
    \textbf{Step 5: Start Constellation Buildup}\;
    \Indp
        Launch constellation satellites\;
    \Indm
    
    \textbf{Step 6: Select Launch Rate Controller and Propagate Initial Period}\;
    \Indp
        \If{launch\_pattern\_type is equilibrium}{
            Apply open-access controller\;
        }
        \ElseIf{launch\_pattern\_type is sat\_feedback}{
            Apply feedback rule to maintain initial populations\;
        }
        Propagate initial period\;
    \Indm
    
    \textbf{Step 7: State Propagation Loop}\;
    \Indp
        \For{each year in model\_horizon}{
            Apply selected fringe launch rate controller and launch constellation satellites\;
            Propagate the orbital environment\;
            \Indp
                \If{model\_type is GMPHD}{
                    Calculate debris distribution over discrete categories as in Step 3\;
                }
        }
    \Indm
    
    \textbf{Step 8: Save Results}\;
    \Indp
        Write final outputs to CSV files\;
    \Indm
    
    \end{algorithm}

\section{Results}
\label{sec:results}

We first validate the model against historical data, then describe the benchmark case, and conduct three exercises to illustrate the types of questions which can be studied using the OPUS framework. These exercises all use the MOCAT-4S propagator. We conclude this section by showing the benchmark case under the GMPHD filter propagator to demonstrate the flexibility of this framework. We start all model projections from the initial population of orbital objects around July 2022.

While the physical parameters of MOCAT-4S have been used to estimate orbital capacity, many of the economic parameters used in the model have yet to be measured. Where possible we have attempted to ensure economic and behavioral parameters are plausible and/or be consistent with historical magnitudes under full compliance. We stress that the results presented below are only intended to demonstrate model capabilities, not to provide specific guidance.

\subsection{Model validation}

To validate OPUS we construct measures of launch patterns by constellations and an open-access fringe. We use historical data on orbital-use patterns compiled in \cite{rao2023preparing}. The dataset is a satellite-year panel covering the period 1957-2022, recording the active lifetime of each satellite launched along with orbital parameters, ownership, sector (i.e. commercial, military, civil, or some combination), country of operator, launch site, launch vehicle, and several other fields. Satellites are removed from the panel once they are estimated to no longer be active, so derelict objects are not included.

We define the constellation population as satellites belonging to the Starlink and OneWeb constellations. We define the open-access fringe population as all other satellites labeled as commercial, and construct a residual ``Other'' category for military and civil government satellites. Note that while the economic theory developed here does not predict the launch patterns of commercial constellations or non-commercial satellites, it does predict that orbital-use patterns of both will affect launch behavior by the commercial fringe via collision risk and any impacts on revenues or costs.\footnote{See \cite{guyot2023satellite} for an economic theory of commercial satellite constellations' orbital-use patterns.} 

\begin{figure}[htbp]
    \centering
    \hspace*{-0.125\textwidth}
    \includegraphics[width=1.25\textwidth]{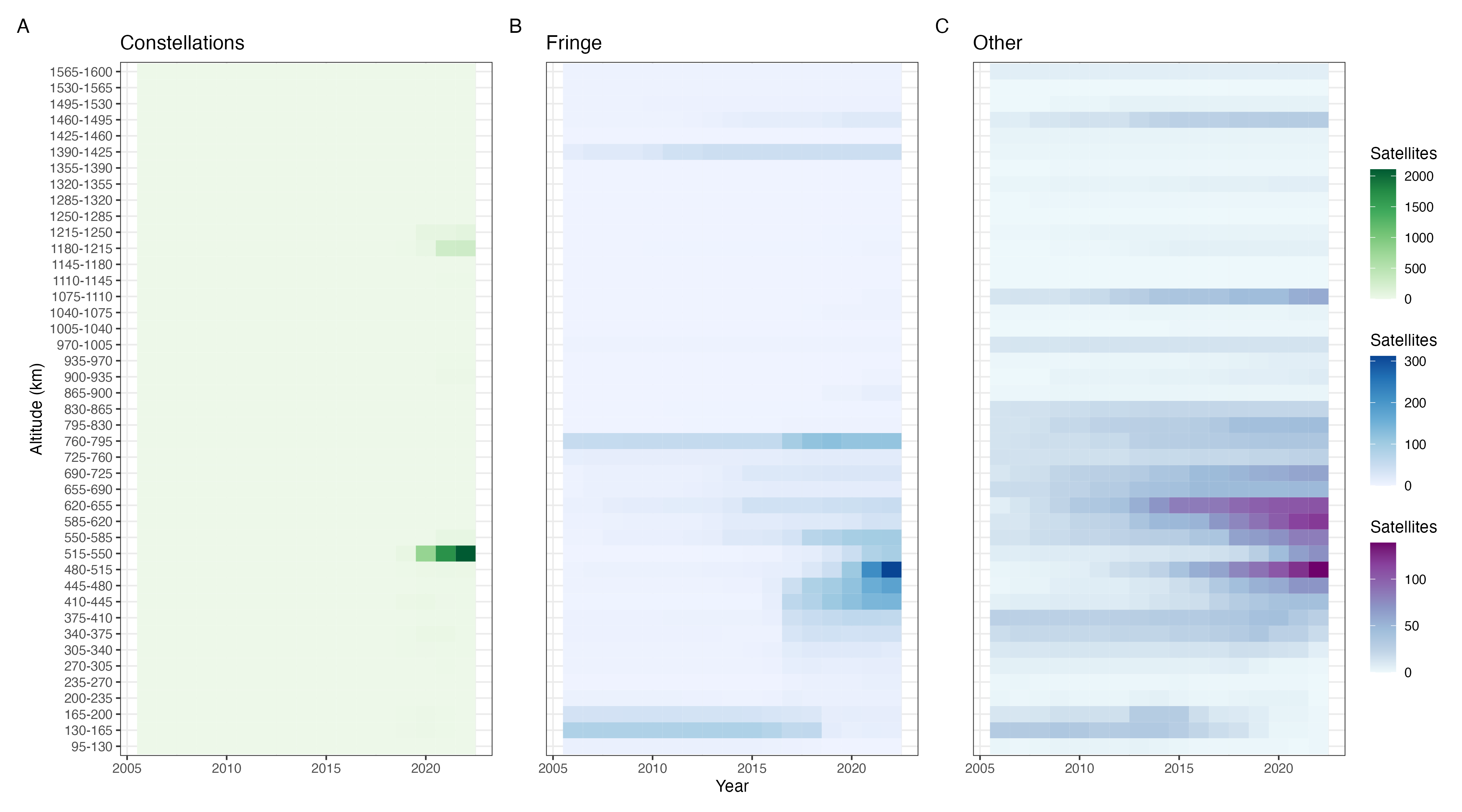}
    \caption{Historical orbital-use patterns over 2005-2022. Panel A shows active constellation satellites, panel B shows active non-constellation commercial satellites, and panel C shows active non-commercial satellites.}
    \label{fig:historical-patterns}
\end{figure}

Figure \ref{fig:historical-patterns} plots the populations of each type of satellite over the 2005-2012 period. We focus on this period since the 25-year deorbit guideline (``25-year rule'') was implemented in 2005, allowing us to compare OPUS predictions under 25-year disposal for the fringe against the historical pattern. Notice that both the constellation and the fringe begin to accumulate appreciable populations only near the end of the sample period, with activity in prior years being dominated by non-commercial actors. The latter half of the sample sees increased activity by non-commercial actors as well. Both fringe and non-commercial satellites appear to avoid altitudes utilized by constellations, most notably in the vicinity of Starlink.

\begin{figure}[htbp]
    \centering
    \hspace*{-0.125\textwidth}
    \includegraphics[width=1.25\textwidth]{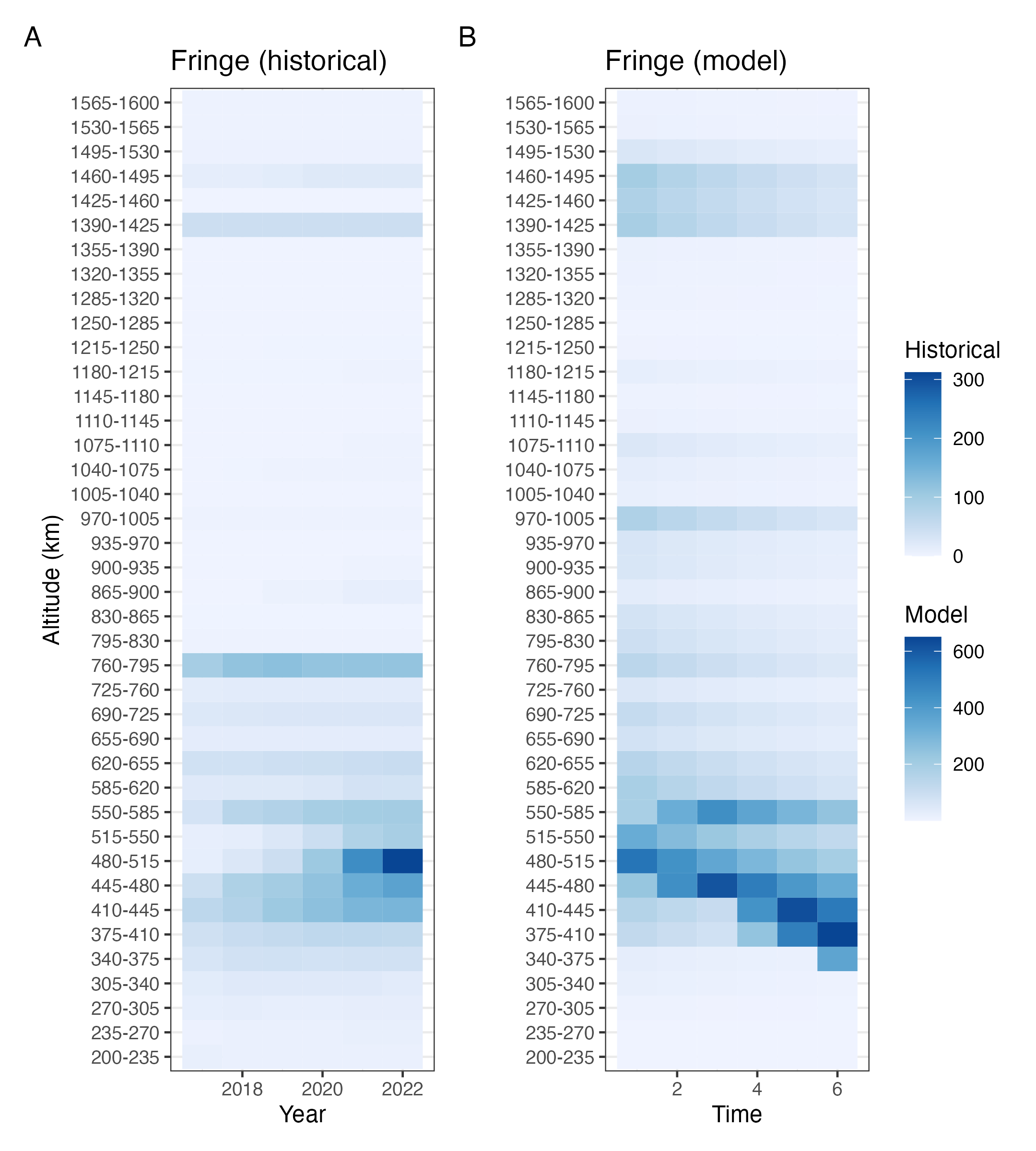}
    \caption{Comparison of historical orbital-use patterns over 2017-2022 against model predictions for first 6 years of a simulation with 25-year disposal and full compliance. Panel A shows active constellation satellites, panel B shows active non-constellation commercial satellites, and panel C shows active non-commercial satellites.}
    \label{fig:model-validation}
\end{figure}

Next, we compare the model's predictions against historical orbital-use patterns over an appropriate sample. Figure \ref{fig:model-validation} plots historical fringe population levels over 2017-2022---when fringe launch activity started picking up---and OPUS predictions over the first 6 years from the initial population. However, there are some caveats in making and interpreting these comparisons.

First, the 2017-2022 period saw substantial declines in launch prices. The vehicle-weighted price index constructed in \cite{corrado2023space} shows prices falling from around 10,000 \$/kg to around 5,000 \$/kg. The increased launch rate may have also contributed to broader sectoral cost reductions, as satellite supply chains benefited from economies of scale and learning-by-doing. These patterns are absent in OPUS predictions, as OPUS currently includes only static values of economic parameters. Future work will extend OPUS to allow for time-varying economic parameters.

Second, key economic parameters---such as the cost of delta-v ($c_{\Delta v}$) and the maximum WTP for fringe service ($\alpha^q_1$)---are not calibrated due to lack of data. These parameters are critical in determining the relative appeal of different orbital locations. The form of the revenue function may be particularly important here. The current linear form, while simple and transparent, means that the cost function is the only purely economic source of variation in the relative attractiveness of different locations. This likely also distorts quantitative comparisons.

Finally, the model begins from the population of orbital objects in July 2022 rather than January 2017. While the initial population can be adjusted, this functionality has not yet been incorporated into OPUS. Additionally, the simulation assumes full compliance with the 25-year rule. though in practice some operators may not comply due to various reasons (e.g. component failure, operating from jurisdictions where local regulators do not mandate 25-year rule compliance).

Despite these caveats, the model appears to capture some important elements of open-access fringe orbit use:
\begin{enumerate}
    \item avoidance of altitudes where the lower and larger constellation operates;
    \item preference for altitudes that are naturally compliant with the 25-year rule (i.e. just below the 585-620 km shell)
    \item among the naturally-compliant altitudes, greater preference for altitudes that are below the nearby constellation rather than above;
    \item avoidance of the lowest shells, where stationkeeping requirements due to drag are highest.\footnote{Although there is a trend in the model predictions to favor lower altitudes over time, we will see in subsequent sections that there appears to be a ``lowest economically-viable altitude''.}
\end{enumerate}

\subsection{Benchmark case}

Next, to illustrate the effect of incorporating economic behavior into propagator projections, we compare the evolution of populations over time---35 years forward from July 2022---under a simple ``satellite feedback'' rule against economic behavior implied by the system of equations \ref{eqn:open-access-condition-system} (``open-access behavior''). Satellite feedback behavior involves launching just enough satellites to maintain the previous period's population levels at all altitudes, ignoring collisions.

\begin{figure}[htbp]
    \centering
    \hspace*{-0.125\textwidth} 
    \includegraphics[width=1.25\textwidth]{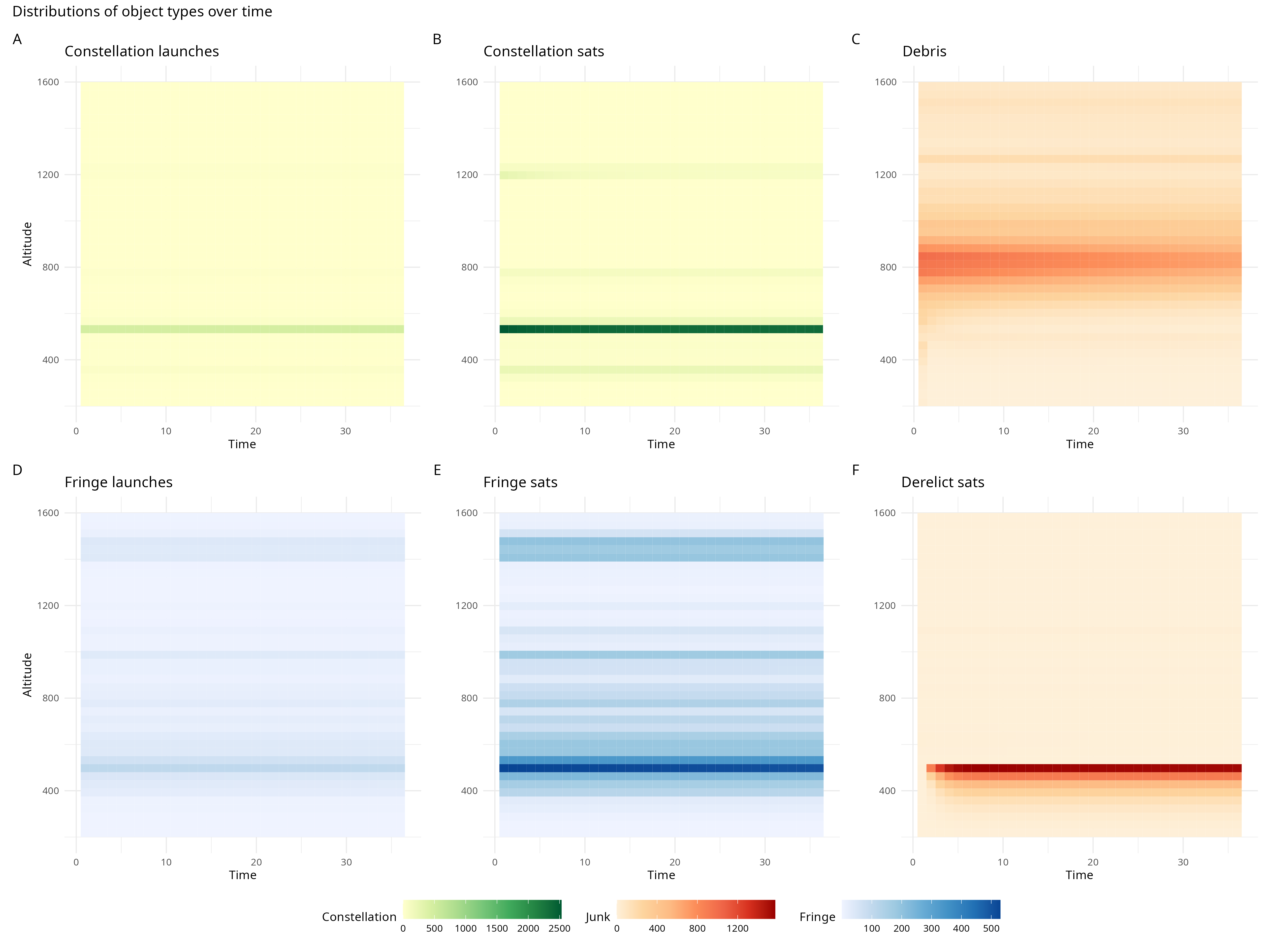}
    \caption{Location distribution of launches and orbital objects over time under satellite feedback behavior.}
    \label{fig:benchmark-MOCAT-satfeedback-35yrs}
\end{figure}

\begin{figure}[htbp]
    \centering
    \hspace*{-0.125\textwidth} 
    \includegraphics[width=1.25\textwidth]{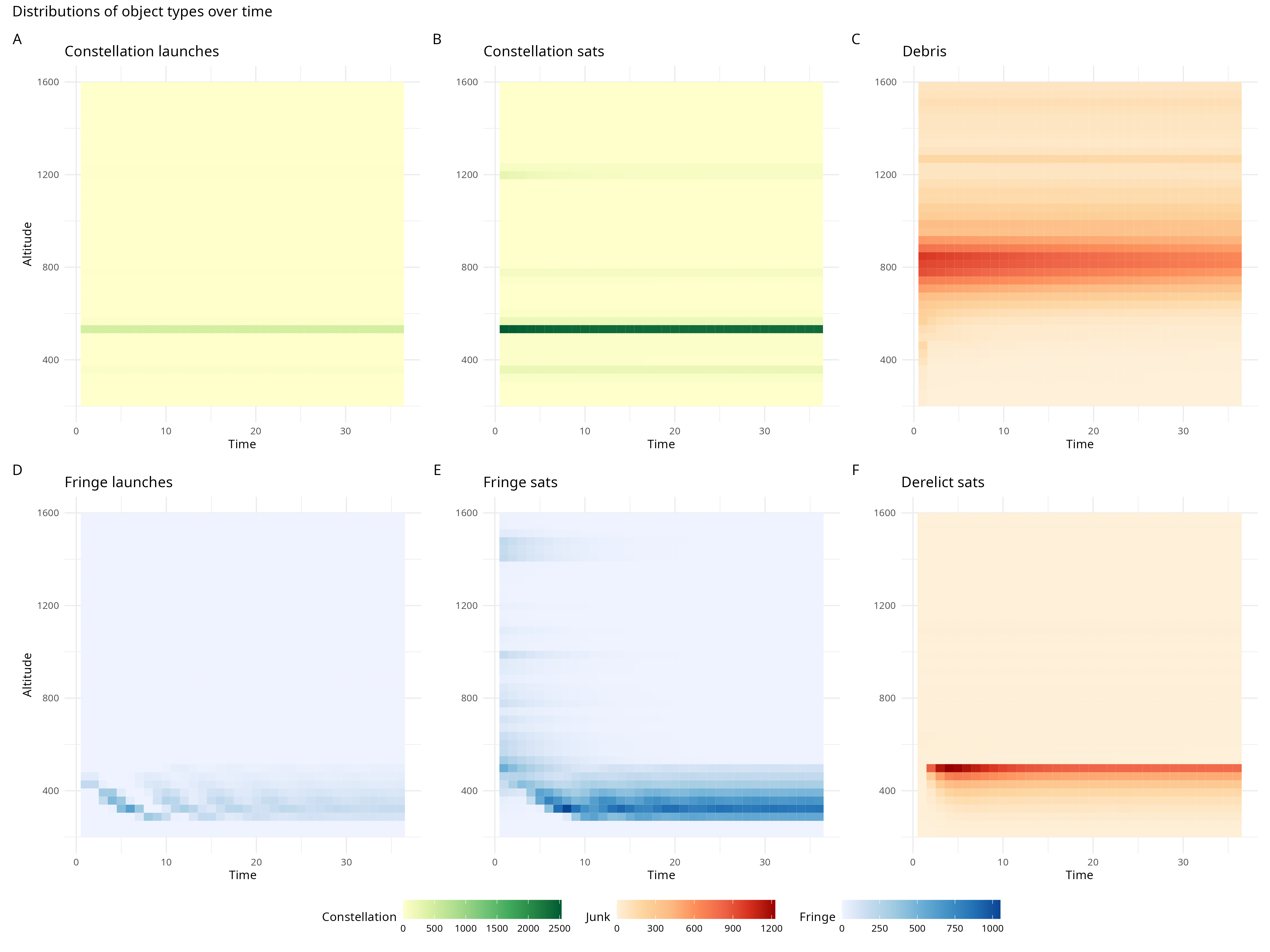}
    \caption{Location distribution of launches and orbital objects over time under open-access behavior.}
    \label{fig:benchmark-MOCAT-equilibrium-35yrs}
\end{figure}

Figure \ref{fig:benchmark-MOCAT-satfeedback-35yrs} plots all 4 populations along with the launch rates for the constellation and fringe under satellite feedback behavior. Figure \ref{fig:benchmark-MOCAT-equilibrium-35yrs} plots the same outcomes under open-access behavior. This case assumes 5-year disposal with full compliance.

There are several notable differences between feedback and open-access behavior. We briefly comment on three particularly striking differences: the periodicity over time and space under equilibrium behavior, the emergence of a lowest economically-viable altitude, and the abandonment of higher altitudes.

\paragraph{Periodicity.} Some form of repetitive behavior often features in debris environment analyses. These are typically either due to solar cycles or the use of autonomous traffic patterns (e.g. \cite{liou2004legend, lewis2008fast})---both exogenously-imposed assumptions. The projections from OPUS instead reflect two economic forces that endogenously respond to the state of the orbital environment:
\begin{enumerate}
    \item As derelicts decay from higher altitudes, it becomes worthwhile to launch to locations with fewer derelicts and lower anticipated future collision risk. That is, open-access behavior implements a type of feedback controller that \emph{to some degree} accounts for the sustainability of particular orbital locations and avoids those that are expected to become less-sustainable in the near future (specifically, the next year). Such behaviors have been documented in both theoretical non-spatial and atheoretical spatial economic models of orbit use \citep{rao2023economics, raoletizia2021}.
    \item Some orbits that are naturally compliant with disposal rules are costlier to operate in than others, due to a combination of the delta-v required to stationkeep and the risk of colliding with existing objects there. As derelicts fill more-valuable orbits, either due to recent launch activity to those orbits or decay from higher orbits, they reduce the overall profitability of using LEO. In addition to spatial variation in where satellites are launched to, this generates variation in the \emph{total} number of satellites launched in each period. This variation also affects the revenues of operating a satellite through market competition, with periods that have fewer satellites leading to greater revenues, spurring further launches.
\end{enumerate}
This endogenous spatiotemporal periodicity is entirely absent in satellite feedback behavior and related exogenous launch models. In the absence of any exogenous shocks the system approaches a steady state, indicating the periodicity eventually fades as the derelicts reach a stationary distribution.

\paragraph{Minimum viable altitude.} As higher orbits that are naturally compliant with the disposal rule fill with constellation satellites or derelict objects, fringe operators move to lower orbits, where debris decays faster. There is a limit to this process of moving lower to avoid debris risk. At a certain altitude the increase in stationkeeping costs outweighs the gain in collision risk reduction, creating a ``floor'' on how low operators choose to go. Indeed, higher naturally-compliant orbits are preferable when they are clean enough to support greater satellite activity. However, since they also retain debris longer, short stretches of higher-altitude usage are matched by longer stretches of lower-altitude usage. The altitude that is most-used for the longest stretch is close to the lowest economically viable altitude. The minimum altitude appears only in satellite feedback behavior to the extent that the initial population already reflects it, and does not adapt to reflect changes in the state of the orbital environment.

\paragraph{Abandonment of higher altitudes.} Figure \ref{fig:cost-fn} shows that the costs of operating increase steeply above the highest naturally-compliant altitude due to the opportunity cost of delta-v expended on disposal (for operators who comply with the rule). Note that the increase in activity at lower altitudes is significant enough to make derelict satellites at higher altitudes difficult to see in the plots, though they will remain there until fragmentation or their eventual decay. To the extent that they comply with disposal rules, fringe operators therefore cease launching to higher altitudes. Satellite feedback behavior actively contradicts this economic response. \\

Finally, before moving on to specific policy exercises, we illustrate the impact of economic behavior on aggregate summary metrics of the orbital environment. Specifically, we compute the Space Sustainability Rating (SSR) index for all objects of a given type across all altitudes \citep{rathnasabapathy2020space}. For this application we calculate the SSR index as the product of the sum of total $p_c$ across all objects and altitudes with the sum of the Environmental Consequence of Breakup (ECOB) index---as the name suggests, a measure of the consequence of a fragmentation on orbit \citep{letiziaetal2017, letiziaetal2018}---across all objects and altitudes. While the SSR index and its components are typically applied to individual missions or objects, we compute the ``aggregate'' index to illustrate the impact of incorporating economic behavior on space sustainability projections. In all cases we normalize the SSR index to one in the initial period. Larger index values indicate less sustainable orbital-use patterns.

\begin{figure}[htbp]
    \centering
    \hspace*{-0.125\textwidth}
    \includegraphics[width=1.25\textwidth]{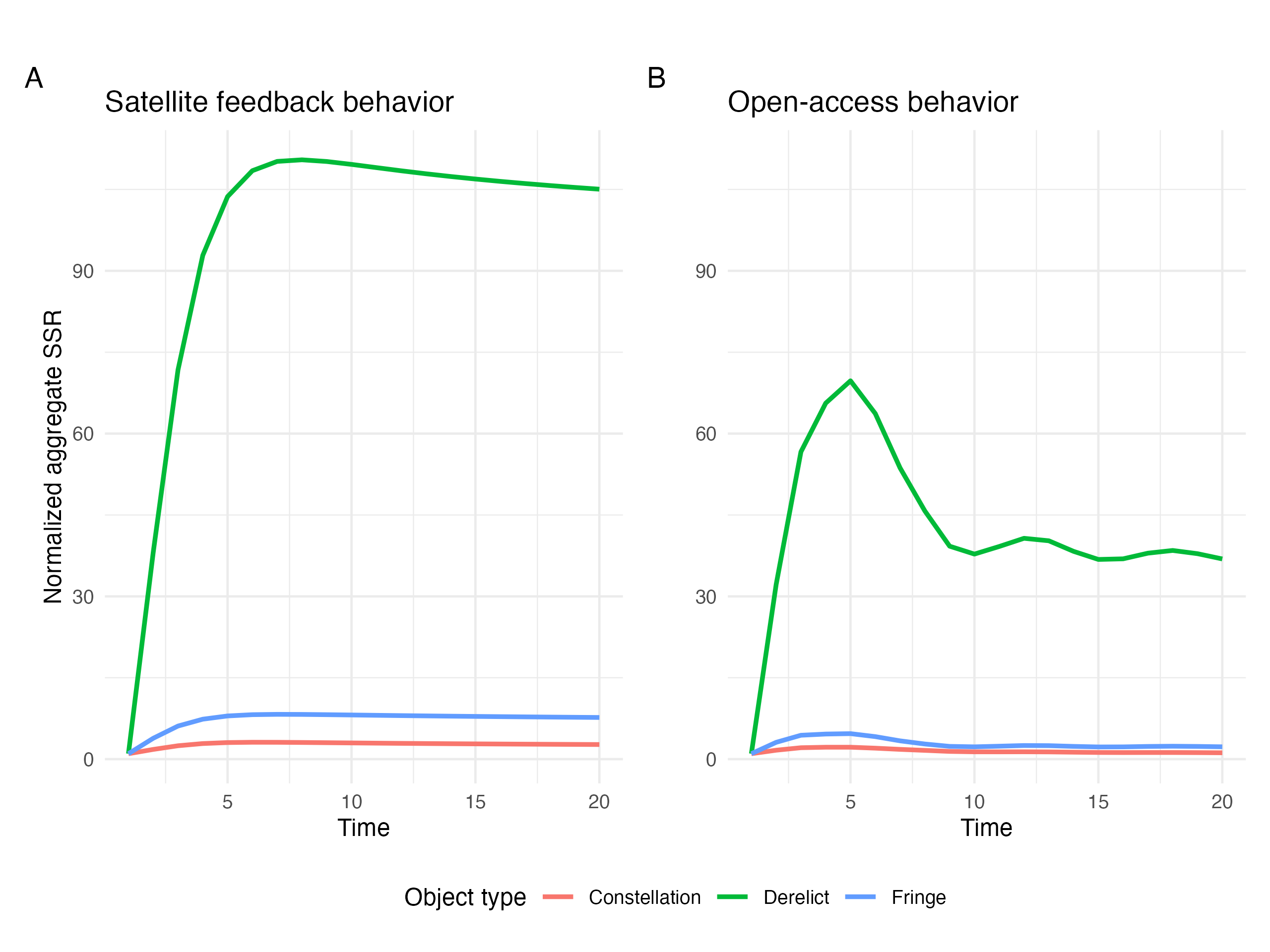}
    \caption{Normalized aggregate SSR index value for all object types under satellite feedback and open-access behavior. Smaller numbers indicate greater sustainability.}
    \label{fig:benchmark-ssr-comparison}
\end{figure}

Figure \ref{fig:benchmark-ssr-comparison} plots the normalized aggregate SSR index by large object types---constellation satellites, fringe satellites, and derelict objects---for the first 20 years of the model simulation. Open-access behavior exhibits different sustainability metric dynamics than satellite feedback behavior. Though the ranking of metrics across object types is generally the same, all object types have lower index values under open-access behavior. Mechanically, the difference lies in the types of feedback controllers the behaviors implement: whereas satellite feedback behavior is a backward-looking controller (launch rates to location $k$ in period $t$ are proportional to the stock in $k$ at $t-1$), the open-access system in equation \ref{eqn:open-access-condition-system} is a forward-looking controller (launch rates to $k$ in $t$ depend on the anticipated stock in $k$ at $t+1$). Intuitively, firms anticipate the returns their asset will generate at each location and choose where to deploy their satellite accordingly.

Both types of behavior feature a sharp increase in the index value for derelict objects as the constellations build up to their target values. For open-access behavior, the initial peak is followed by damped oscillations as fringe operators adjust from the initial condition toward a stationary distribution.\footnote{Note that a stationary distribution need not be reached. Firms may anticipate runaway debris growth as an open-access equilibrium and rationally choose to launch such that it occurs \citep{rao2023economics}. For the simulations in this paper we set economic parameters to prevent this outcome.} By contrast, satellite feedback behavior reaches a stationary distribution much more quickly---the only source of significant changes from the initial conditions are the constellations, particularly the lower and larger one.

OPUS also enables calculation of economic metrics, like the annual expected maximum economic welfare from orbit-use. The annual expected maximum economic welfare, $\lambda_{t}$, is a dollar metric of the annual social value of orbit use under a given policy regime. Like the SSR, it has a ``probability $\times$ consequence'' form with the probability being $1-p_c$, i.e. the survival (rather than collision) probability. Unlike the SSR, the ``consequence'' is the net present value of all fringe satellites. Formally:

\begin{equation}
    \lambda_{t} = \sum_{k \in K} (1 - P_{2k}(S_{\cdot \cdot t},D_{\cdot \cdot t})) \times S_{2 k t} \left( \sum_{\Delta t = 0}^{\mu^{-1}} \frac{q_k(S_{2 \cdot t})}{(1+r)^{\Delta t}} - c_k \right).
\end{equation}

Economically, $\lambda_{t}$ can be interpreted as the expectation of the upper envelope of lifetime social welfare (i.e. social value generated net of real resource costs) generated by the open-access fringe in year $t$, assuming all satellites were just launched and expected to survive their full design lifetimes if they survive year $t$.\footnote{Note the order of terms: ``expected maximum'' rather than ``maximum expected''. The expectation is taken over maximal values assuming no collisions after year $t$, rather than maximizing an expected value that incorporates collision risk in every operational year.} This calculation is simpler than the full expected net present value of social welfare from the satellite fleet used in other economic analyses, e.g. \cite{rao2020orbital}; since our goal is only to demonstrate the type of metrics that can be obtained using OPUS, we do not implement the full net present value calculation.\footnote{While any taxes applied would reduce the firm's profits, profits are only a component of the social welfare generated from orbit use. The tax revenue collected still contributes to economic welfare, and may even be used to fund public goods such as debris remediation. Subtracting taxes form $\lambda_t$ would convert the metric from an upper bound on social welfare to an upper bound on profits.} In contrast to the SSR index, larger expected maximum values indicate more socially-valuable orbit use.\footnote{The expected maximum welfare is non-monotone in the number of satellites in orbit. At low levels, increasing the number of satellites will increase the expected value even as collisions become more frequent. At high levels, the gain in expected value from additional satellites is offset by both the reduction in their survival probability (due to collisions) and the reduction in their individual value (due to competition).} Note that open-access behavior will not maximize economic welfare without natural capital pricing policies like optimal Pigouvian taxes to address the externalities between satellite operators.

\begin{figure}[htbp]
    \centering
    \hspace*{-0.125\textwidth}
    \includegraphics[width=1.25\textwidth]{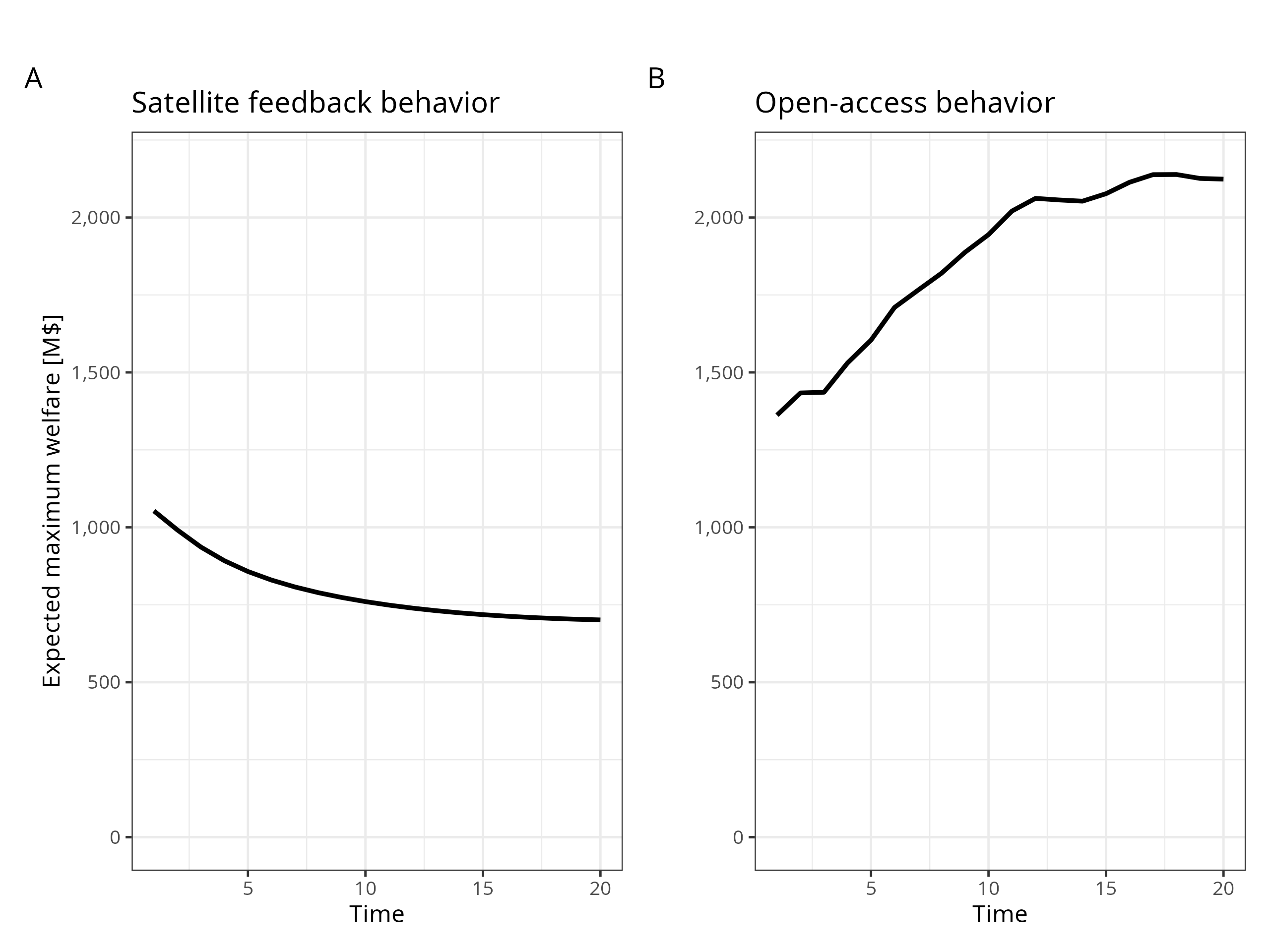}
    \caption{Annual expected maximum economic welfare under satellite feedback and open-access behavior. Larger numbers indicate greater social value from orbit use.}
    \label{fig:benchmark-welfare-comparison}
\end{figure}

\begin{figure}[htbp]
    \centering
    \hspace*{-0.125\textwidth}
    \includegraphics[width=1.25\textwidth]{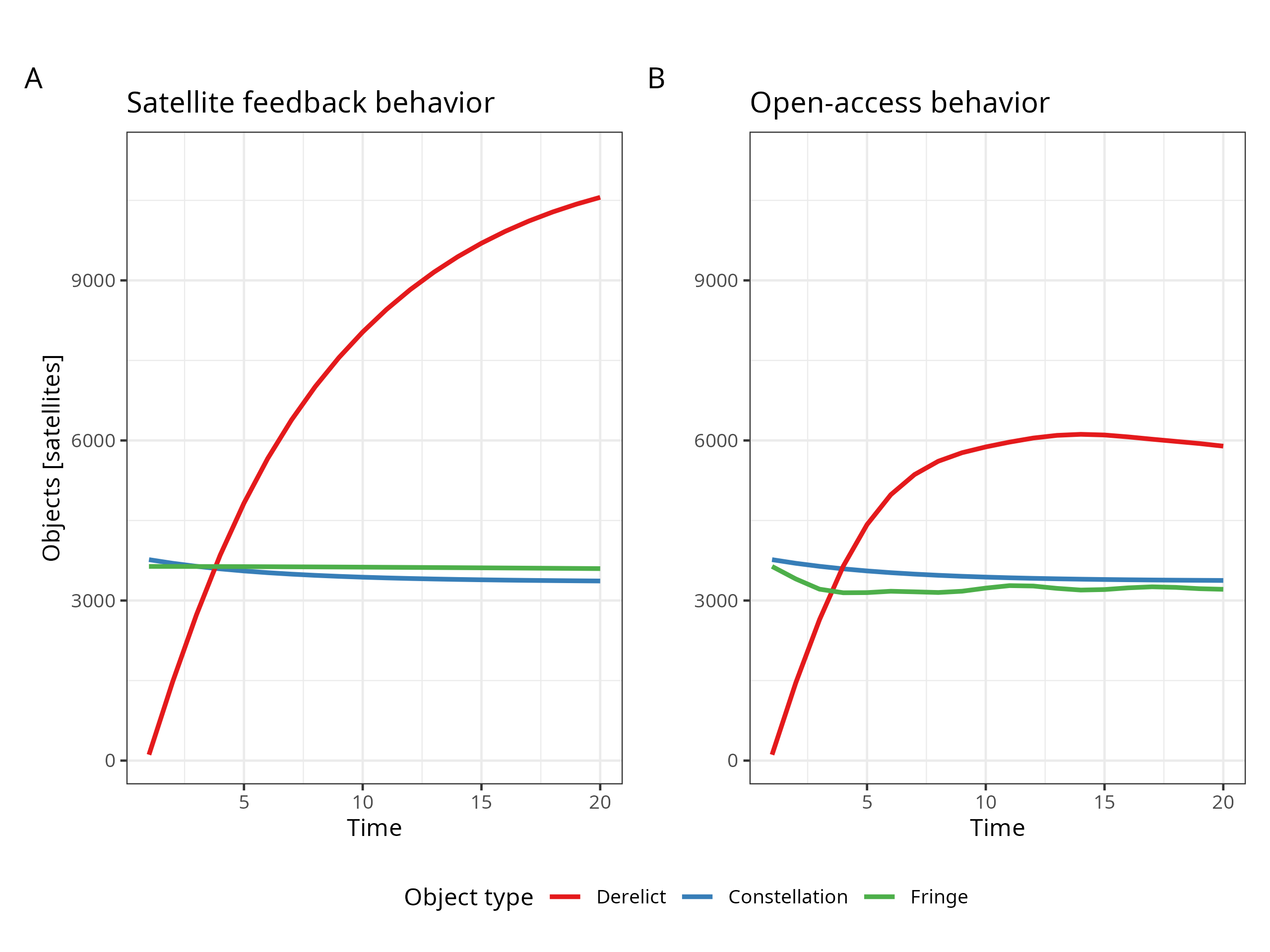}
    \caption{Total object accumulations under satellite feedback and open-access behavior, grouped by object type.}
    \label{fig:benchmark-total-sats-comparison}
\end{figure}

\subsection{25- vs 5-year disposal rules}
\label{sec:25-vs-5-year-rules}

Having validated the model and established some properties of open-access launching behavior, we turn to our first policy evaluation exercise: comparing the effects of 5-year and 25-year disposal rules. We assume for this exercise that binding international disposal rules are implemented with full compliance. To assess the policies we evaluate the patterns of fringe satellite and derelict accumulation, the normalized aggregate SSR index, and the annual expected maximum welfare from orbit-use accrued under each policy. We emphasize once again that these exercises are only demonstrative of the framework's capabilities; while relative comparisons and qualitative patterns can provide some insight into the mechanics of open access, detailed quantitative insights cannot yet be drawn from model results.

\begin{figure}[htbp]
    \centering
    \hspace*{-0.125\textwidth}
    \includegraphics[width=1.25\textwidth]{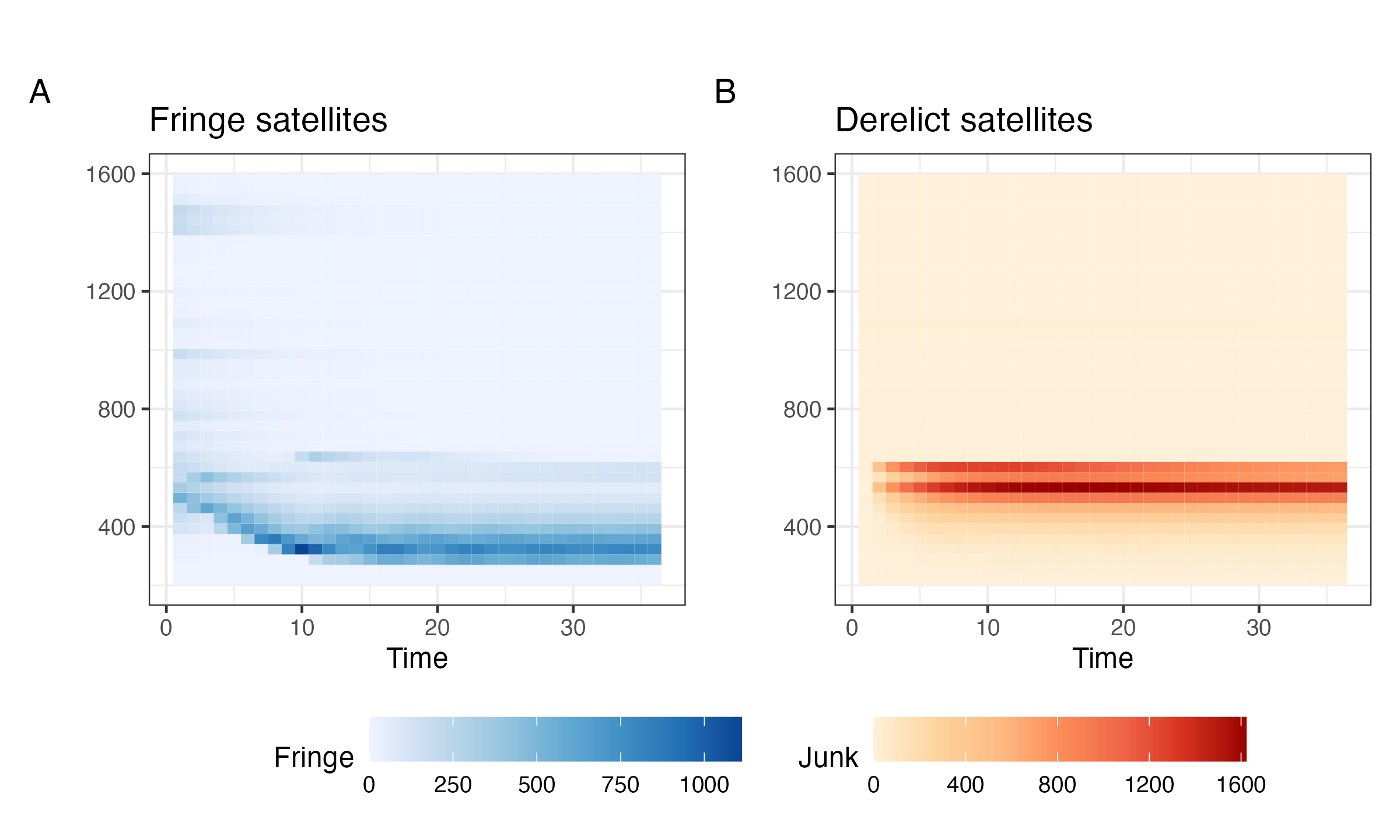}
    \caption{Accumulation of fringe satellites and derelict objects under 25-year disposal rule with full compliance.}
    \label{fig:25-year-panel}
\end{figure}

\begin{figure}[htbp]
    \centering
    \hspace*{-0.125\textwidth}
    \includegraphics[width=1.25\textwidth]{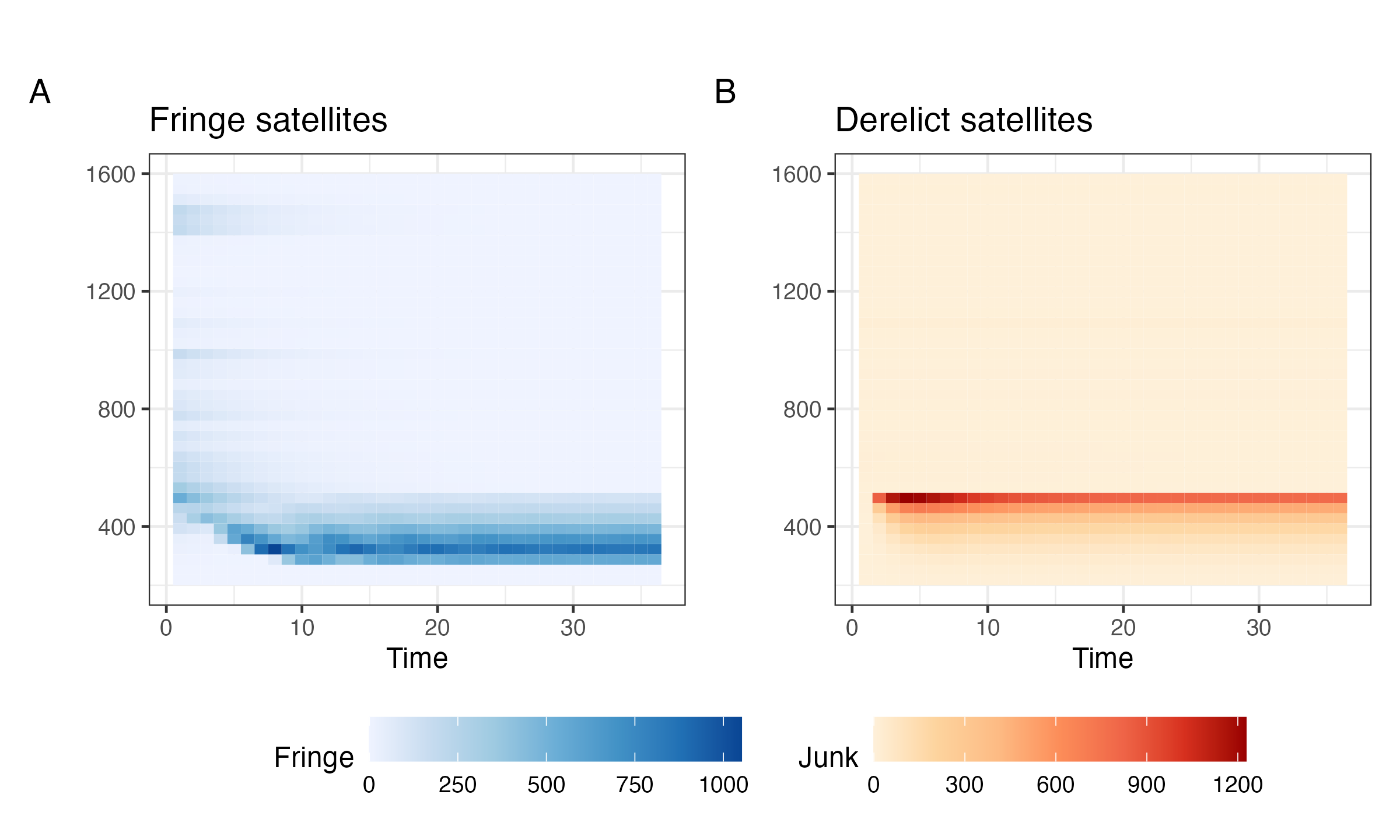}
    \caption{Accumulation of fringe satellites and derelict objects under 5-year disposal rule with full compliance.}
    \label{fig:5-year-panel}
\end{figure}

Figures \ref{fig:25-year-panel} and \ref{fig:5-year-panel} show the patterns of fringe satellite and derelict object accumulation over 35-year horizons from the initial condition. Under both disposal rules fringe satellites avoid the location where the lower and larger constellation is, but under the 25-year rule fringe satellites spread out over more altitudes and use the lower altitudes less intensively. 

Under the 25-year rule two derelict ``hotspots'' form, one associated with the lower and larger constellation and one associated with the (comparatively fewer) fringe satellites that locate above the constellation. The higher derelict hotspot due to the fringe satellites fades over time as fringe satellites largely abandon the higher reaches of the naturally-compliant region after initial periods of low usage.

Under the 5-year rule satellites cluster more tightly below the lower and larger constellation, with no accumulation above the constellation. Consequently the only derelict hotspot is due to the constellation. The peak fringe satellite accumulation under the 5-year rule is lower than under the 25-year rule, as the concentration of satellites in a smaller region increases the congestion there.

\begin{figure}[htbp]
    \centering
    \hspace*{-0.125\textwidth}
    \includegraphics[width=1.25\textwidth]{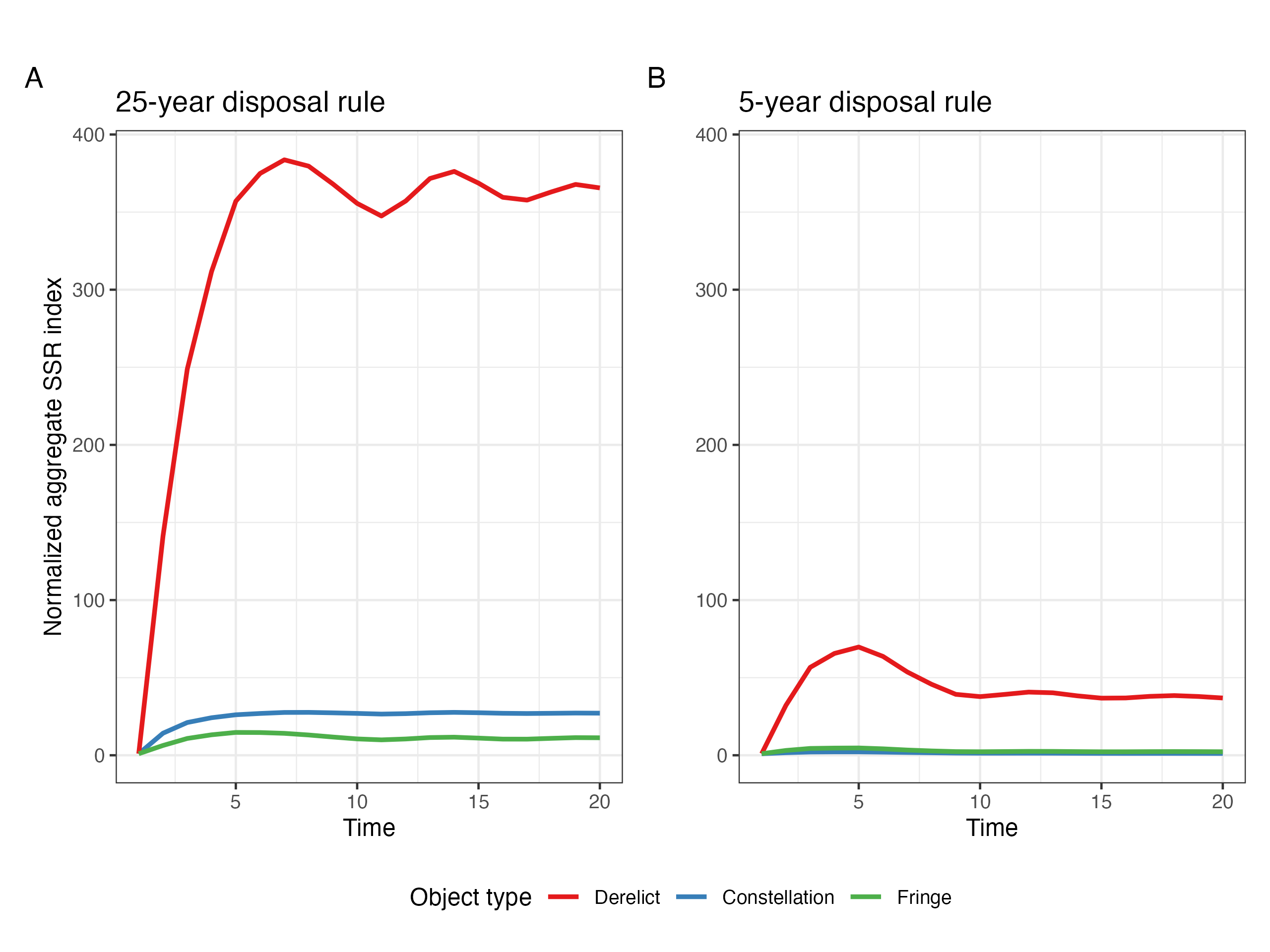}
    \caption{Normalized aggregate SSR index values for all object types under open-access behavior given full compliance with 25- or 5-year disposal rules.}
    \label{fig:25-vs-5-ssr-comparison}
\end{figure}

Figure \ref{fig:25-vs-5-ssr-comparison} shows the normalized aggregate SSR index for the disposal rules over a 20-year horizon following the initial condition. As expected, 5-year disposal results in a substantially lower normalized aggregate SSR index compared to 25-year disposal. Most of the improvement is driven by reduction in the stock of derelicts on orbit, with smaller improvements due to reduced risks facing fringe and constellation satellites.

\begin{figure}[htbp]
    \centering
    \hspace*{-0.125\textwidth}
    \includegraphics[width=1.25\textwidth]{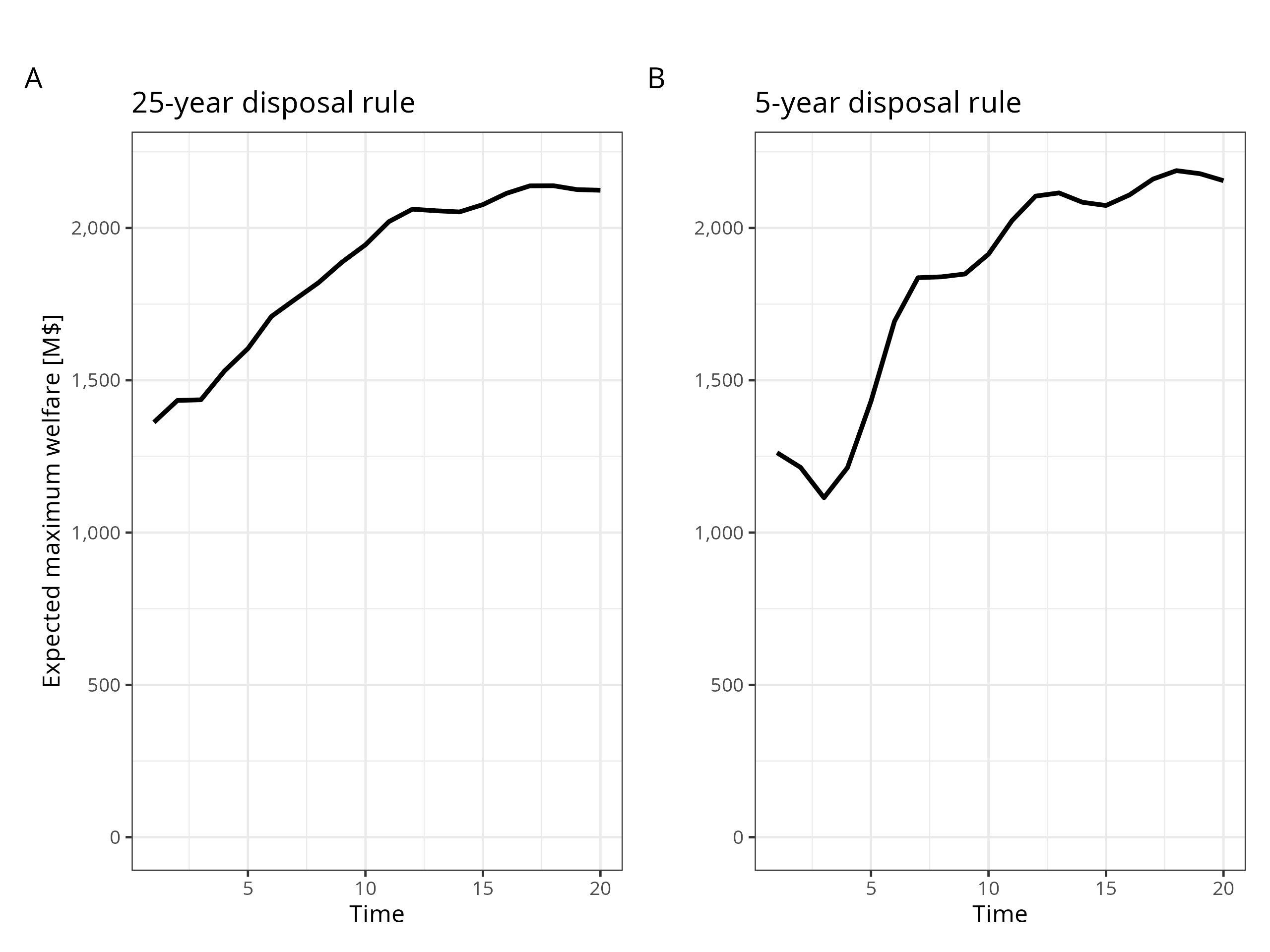}
    \caption{Expected maximum economic welfare from fringe satellites under open-access behavior given full compliance with 25- or 5-year disposal rules.}
    \label{fig:25-vs-5-emw-comparison}
\end{figure}

Figure \ref{fig:25-vs-5-emw-comparison} shows the expected maximum economic welfare under both disposal rules. Unlike the SSR index value, the comparison is less clear. The 25-year rule allows for greater initial economic welfare, since more fringe satellites can consistently be maintained by spreading out over more altitudes. In contrast, the 5-year disposal rule leads to an initial decline in economic welfare as the cost of compliance forces operators to cluster at lower altitudes. This induces some operators to refrain from launching (relative to the 25-year rule counterfactual), reducing economic welfare. By the end of the simulation horizon, the improvements in the orbital environment lead to slightly higher economic welfare under the 5-year rule.

\begin{figure}[htbp]
    \centering
    \hspace*{-0.125\textwidth}
    \includegraphics[width=1.25\textwidth]{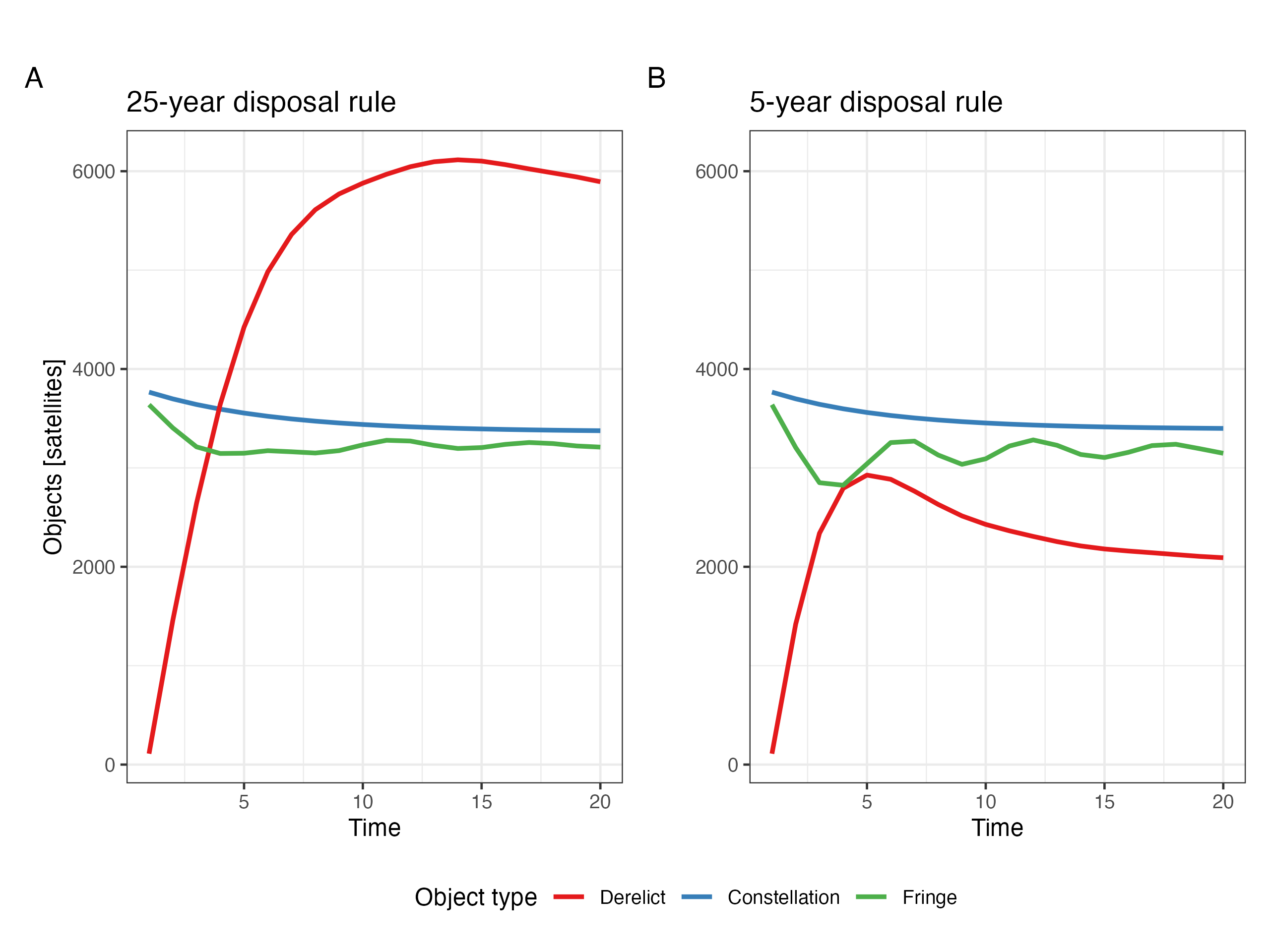}
    \caption{Total numbers of constellation, fringe, and derelict satellites under open-access behavior given full compliance with 25- or 5-year disposal rules.}
    \label{fig:25-vs-5-totals-comparison}
\end{figure}

Figure \ref{fig:25-vs-5-totals-comparison} plots the total numbers of constellation, fringe, and derelict satellites to explore the drivers of the outcomes in Figures \ref{fig:25-ouf-vs-5-ssr-comparison} and \ref{fig:25-vs-5-emw-comparison}. The difference in aggregate SSR for derelict objects appears to be due in large part to differing numbers of derelict objects. As expected, the 25-year disposal rule allows for a substantially larger buildup of derelicts than the 5-year rule. The total number of constellation satellites decrease over time as the initial condition includes constellation (i.e. ``slotted'') satellites in locations other than the two being simulated, while the two being simulated are larger than the extant population in the initial condition (generating concentrated buildups of derelicts). Perhaps unexpectedly, the 5-year disposal rule also leads to fewer fringe satellites than the 25-year disposal rule, with more oscillatory behavior. This is consistent with the patterns observed in Figures \ref{fig:25-year-panel} and \ref{fig:5-year-panel}. The 5-year rule forces operators to cluster in a tigher range of altitudes, leaving fewer locations that are economically valuable and increasing their sensitivity to fluctations in the distributions of derelict and other fringe satellites.

\subsection{Orbital-use fees proportional to anticipated $p_c$}
\label{sec:ouf-prop-to-pc}

A unique feature of IAMs like OPUS is the ability to compare non-economic policies like binding disposal rules and economic policies like orbital-use fees (OUFs) in an internally-consistent manner. While dynamically-optimal OUFs as in \citep{rao2020orbital} are computationally expensive to calculate in even the single-location case, simpler implementations may still be worth exploring.\footnote{Dynamically-optimal OUFs can be calculated in future iterations of OPUS.} We therefore consider an exercise using OUFs that are proportional to the anticipated next-period aggregate $p_c$, i.e. setting $\tau_{kt} = \varepsilon P_{2k}(S_{\cdot \cdot t}, D_{\cdot \cdot t})$ for a constant value of $\varepsilon$ in equation \ref{eqn:open-access-condition-system}. We arbitrarily set $\varepsilon = 0.5$ for this exercise. This value implies operators at location $k$ are charged an annual fee equal to 50\% of the expected replacement cost of a satellite deployed to $k$.\footnote{To see this, note that equation \ref{eqn:open-access-condition-system} can be converted from rate units to dollar units by multiplying through by $c_k$, making the tax level equal to $\tau_{kt} c_k = 0.5 P_{2k}(S_{\cdot \cdot t}, D_{\cdot \cdot t}) c_k$. See \cite{rao2023economics} for more on this point.}

This type of simple OUF only strengthens the already-extant incentive for fringe operators to control future collision risk growth. It does not directly introduce incentives to account for additional consequences such as runaway debris growth or consequences to other operators in other shells, except insofar as those consequences are correlated with $p_c$ in the location the operator is considering launching to next period. As before, we evaluate the policies by the patterns of fringe satellite and derelict object accumulation, the normalized aggregate SSR index, and the expected maximum economic welfare under each policy.

\begin{figure}[htbp]
    \centering
    \hspace*{-0.125\textwidth}
    \includegraphics[width=1.25\textwidth]{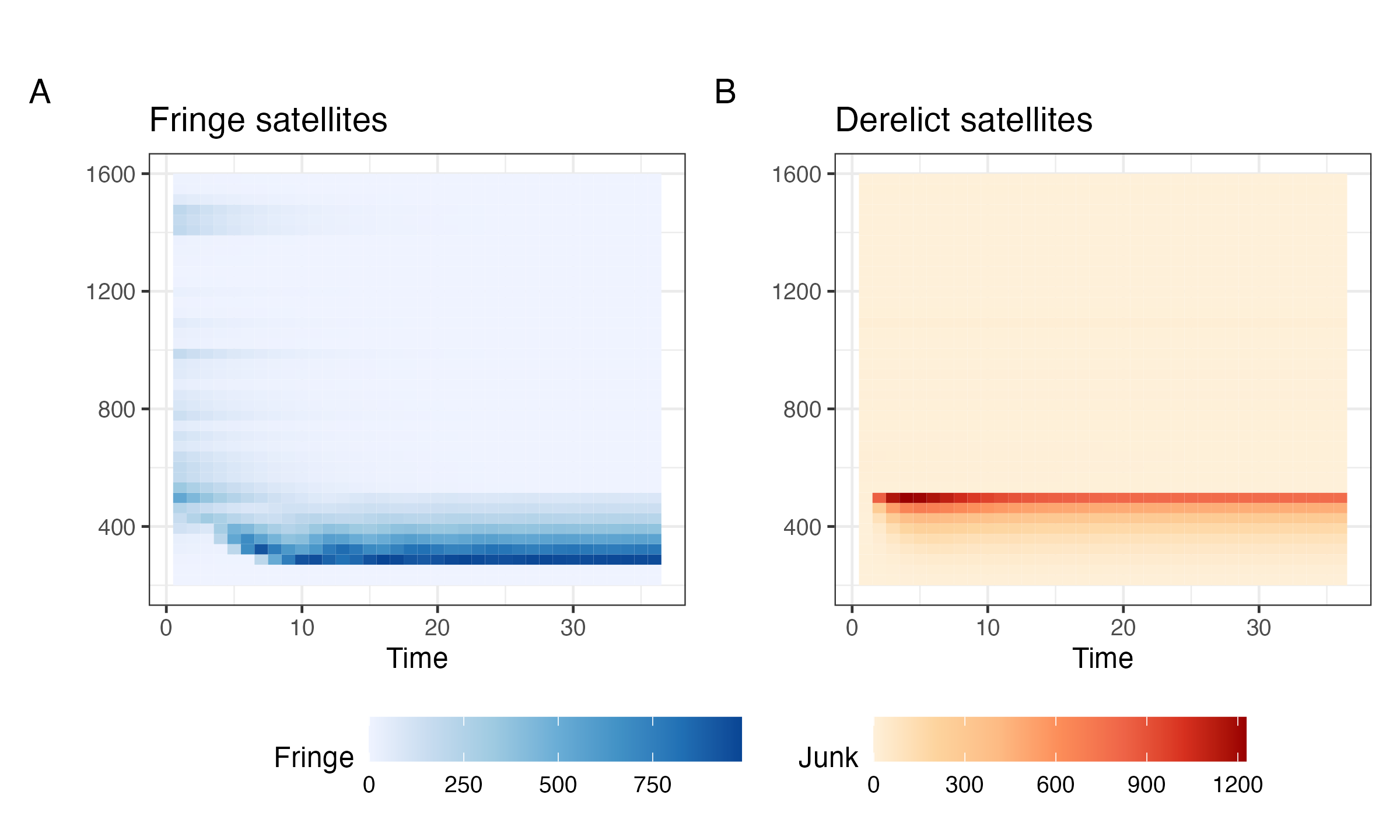}
    \caption{Accumulation of fringe satellites and derelict objects under 25-year disposal rule with full compliance and a 50\%-anticipated-$p_c$ OUF.}
    \label{fig:25-year-ouf-panel}
\end{figure}

Figure \ref{fig:25-year-ouf-panel} shows the patterns of fringe satellite and derelict object accumulation. The 50\%-anticipated-$p_c$ OUF has a noticeable effect on accumulation patterns. The second hotspot above the constellation visible in Figure \ref{fig:25-year-panel} no longer emerges, and the fringe satellites cluster below the constellation even more strongly than under the 5-year disposal rule (seen in Figure \ref{fig:5-year-panel}). Indeed, with the OUF in place fringe operators strongly cluster near the minimum economically viable orbit, further reducing derelict accumulation relative to even the 5-year disposal rule.

\begin{figure}[htbp]
    \centering
    \hspace*{-0.125\textwidth}
    \includegraphics[width=1.25\textwidth]{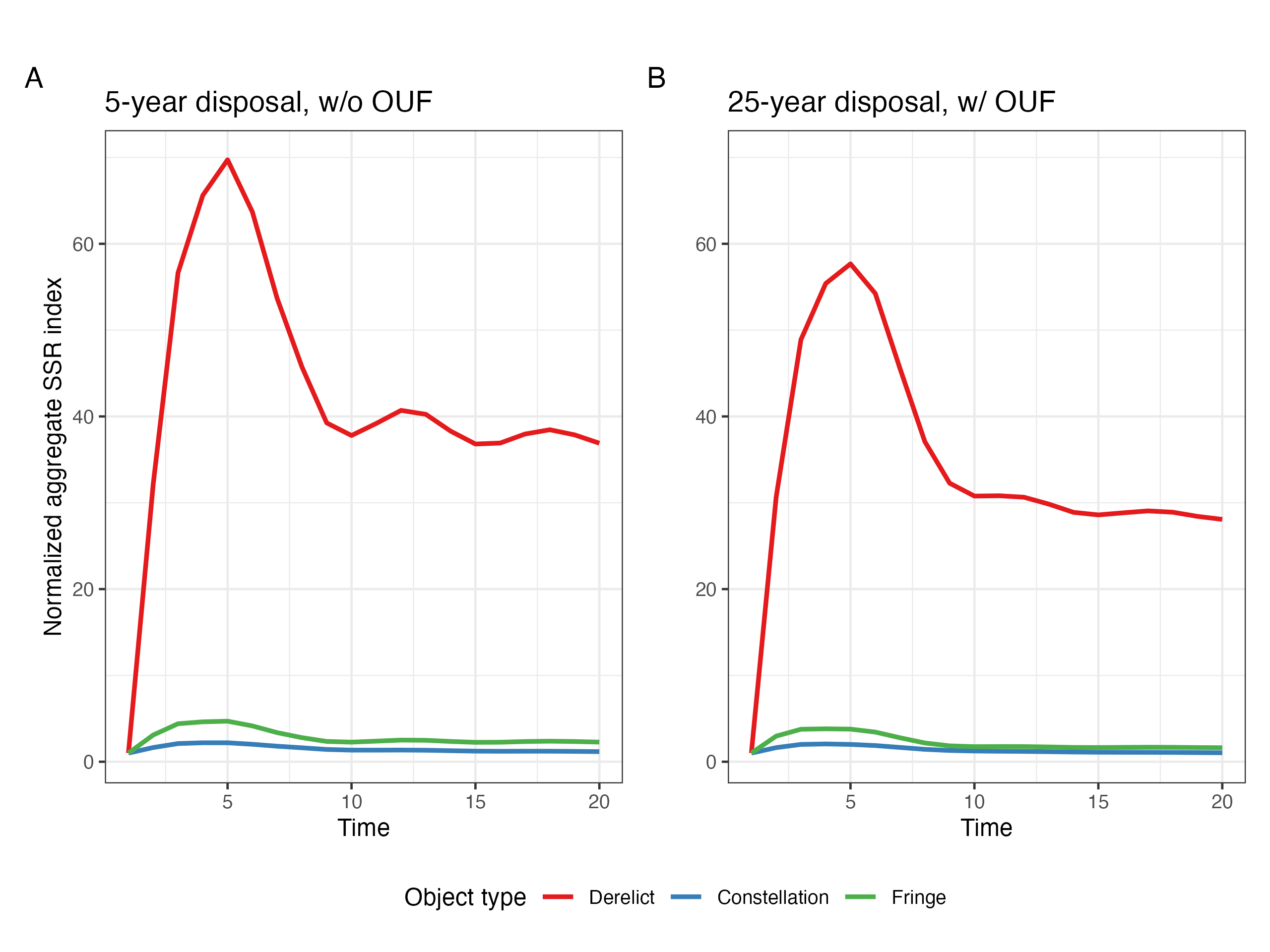}
    \caption{Normalized aggregate SSR index values for all object types under open-access behavior given full compliance with 25- or 5-year disposal rules.}
    \label{fig:25-ouf-vs-5-ssr-comparison}
\end{figure}

\begin{figure}[htbp]
    \centering
    \hspace*{-0.125\textwidth}
    \includegraphics[width=1.25\textwidth]{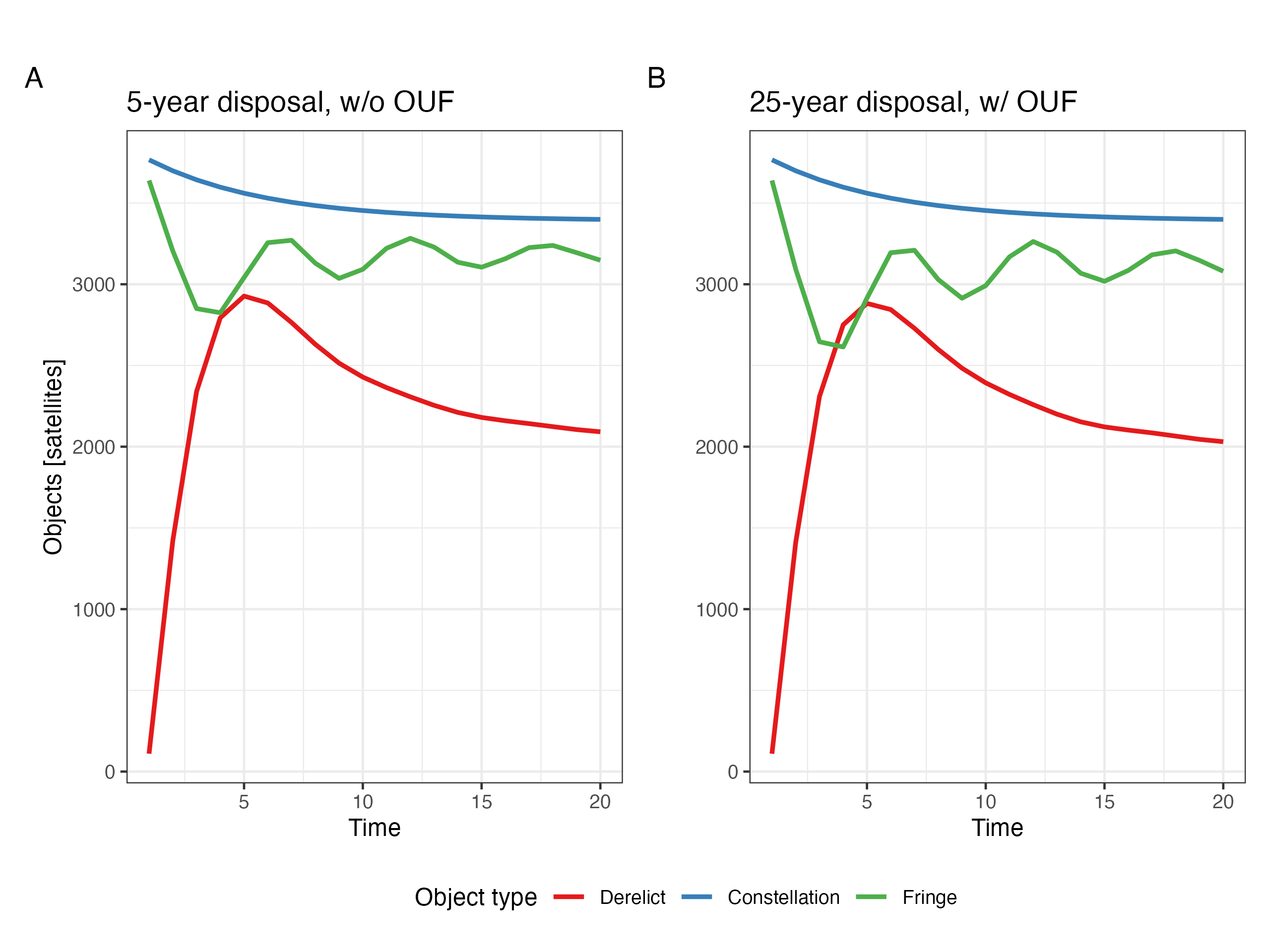}
    \caption{Total numbers of constellation, fringe, and derelict satellites under open-access behavior given full compliance with 25- or 5-year disposal rules.}
    \label{fig:25-ouf-vs-5-totals-comparison}
\end{figure}

Figure \ref{fig:25-ouf-vs-5-ssr-comparison} compares the normalized aggregate SSR index for 5-year disposal without the OUF against 25-year disposal with the OUF. Perhaps surprisingly---but consistent with the accumulation patterns in Figure \ref{fig:25-year-ouf-panel}---25-year disposal with the OUF results in lower SSR index values for the derelict population than the 5-year disposal rule. That this is largely attributable to the pattern of satellite and derelict accumulation rather than their total numbers can be seen from Figure \ref{fig:25-ouf-vs-5-totals-comparison}, which shows the difference in totals is smaller than the difference in SSR index values.

\begin{figure}[htbp]
    \centering
    \hspace*{-0.125\textwidth}
    \includegraphics[width=1.25\textwidth]{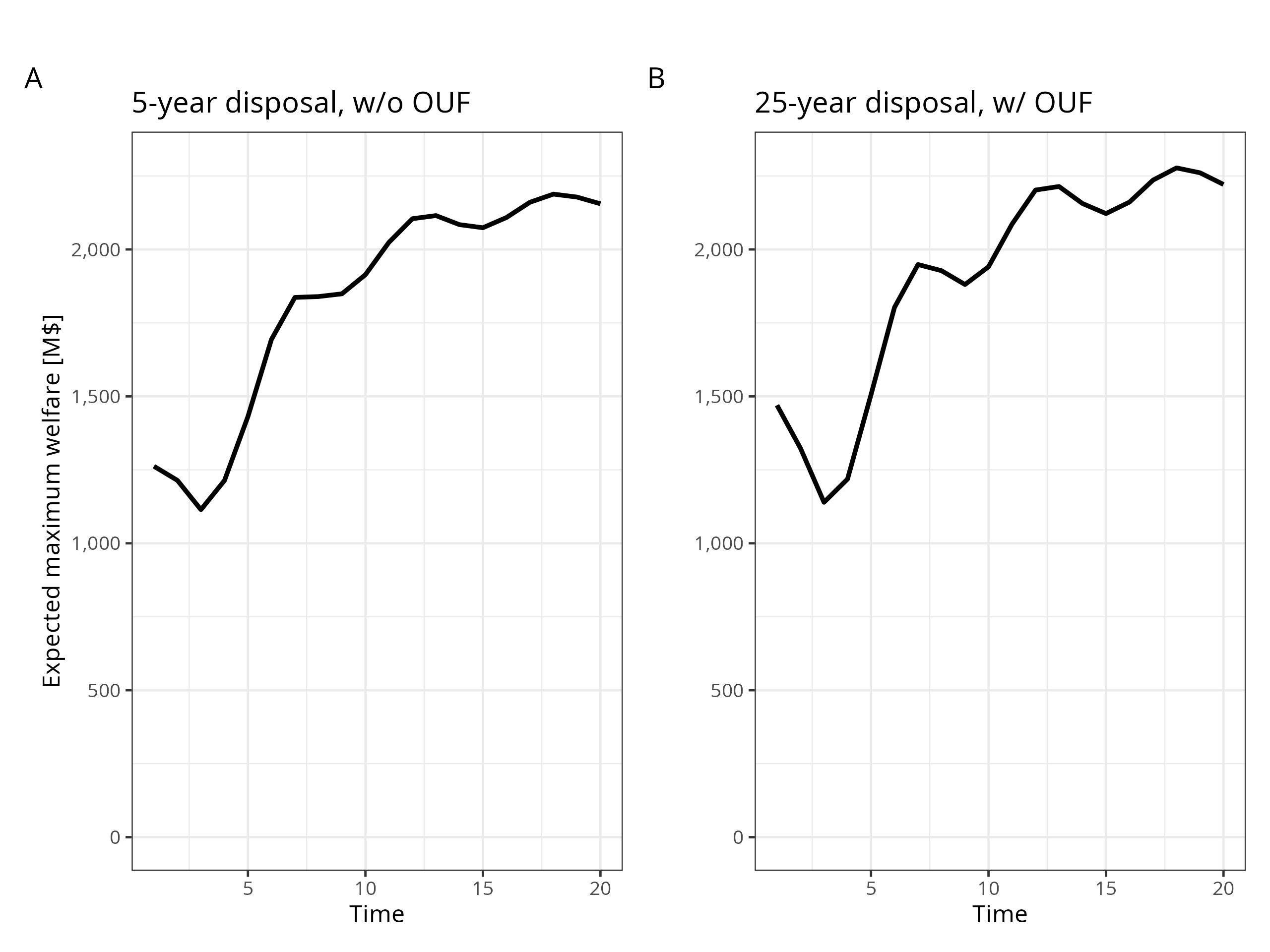}
    \caption{Expected maximum economic welfare for fringe satellites under open-access behavior given full compliance with 5-year disposal rule or 25-year disposal rule with a location-specific 50\%-anticipated-$p_c$ OUF.}
    \label{fig:25-ouf-vs-5-emw-comparison}
\end{figure}

Figure \ref{fig:25-ouf-vs-5-emw-comparison} compares the expected maximum economic welfare under both policies. The expected maximum welfare similar in both cases, though the OUF starts at a higher level, ends at a higher level, and features stronger oscillations. These features are intuitive from the nature and structure of the OUF. Initially, when higher altitudes are clearer, operators can use them, allowing for more satellites and lower economic costs overall. As those orbits fill and anticipated $p_c$ from continuing to launch there grows, operators choose to cluster at lower altitudes to avoid fee liability. Since the OUF moves pro-cyclically with object accumulation and anticipated $p_c$, the oscillations due to debris dynamics are relatively amplified compared to the 5-year rule (though still damped). The 5-year rule does not face operators with this kind of dynamic incentive to keep the orbit usable and also does not allow them to use higher orbits while they are clear, reducing economic welfare generated.

Finally, Figure \ref{fig:25-ouf-vs-5-totals-comparison} compares total object accumulations under both policies. The overall numbers of derelicts are fairly similar though the OUF results in slightly smaller numbers. The initial trough in fringe satellite accumulation is deeper, though as time progresses both feature similar oscillations and approach similar levels.

Again, we emphasize that these results are demonstrative of model capabilities and the quantitative magnitudes should not be interpreted as specific predictions. However, the qualitative patterns point to an important observation in environmental and public economics: binding policies that change the behavior of rational actors function as taxes, whether implemented as such or not \citep{baumol1988theory, fullerton2001framework, bovenberg2002environmental}. Unlike the implicit taxes created by command-and-control policies such as deorbit mandates, explicit taxes allow operators to identify productive margins of substitution and can be used to directly target a desired policy outcome or metric \citep{tietenberg2013reflections}. They can also raise revenue, which can be used toward other policy goals (e.g. financing public goods such as debris remediation, or cutting other taxes) \citep{goulder1995environmental}. Yet the incidence of the policies, and therefore their political economies, can be quite different \citep{fullerton2011six}. Further research can help elucidate these differences in the orbital context and inform policy design.

\subsection{The GMPHD propagator}

\begin{figure}[htbp]
    \centering
    \hspace*{-0.125\textwidth}
    \includegraphics[width=\textwidth]{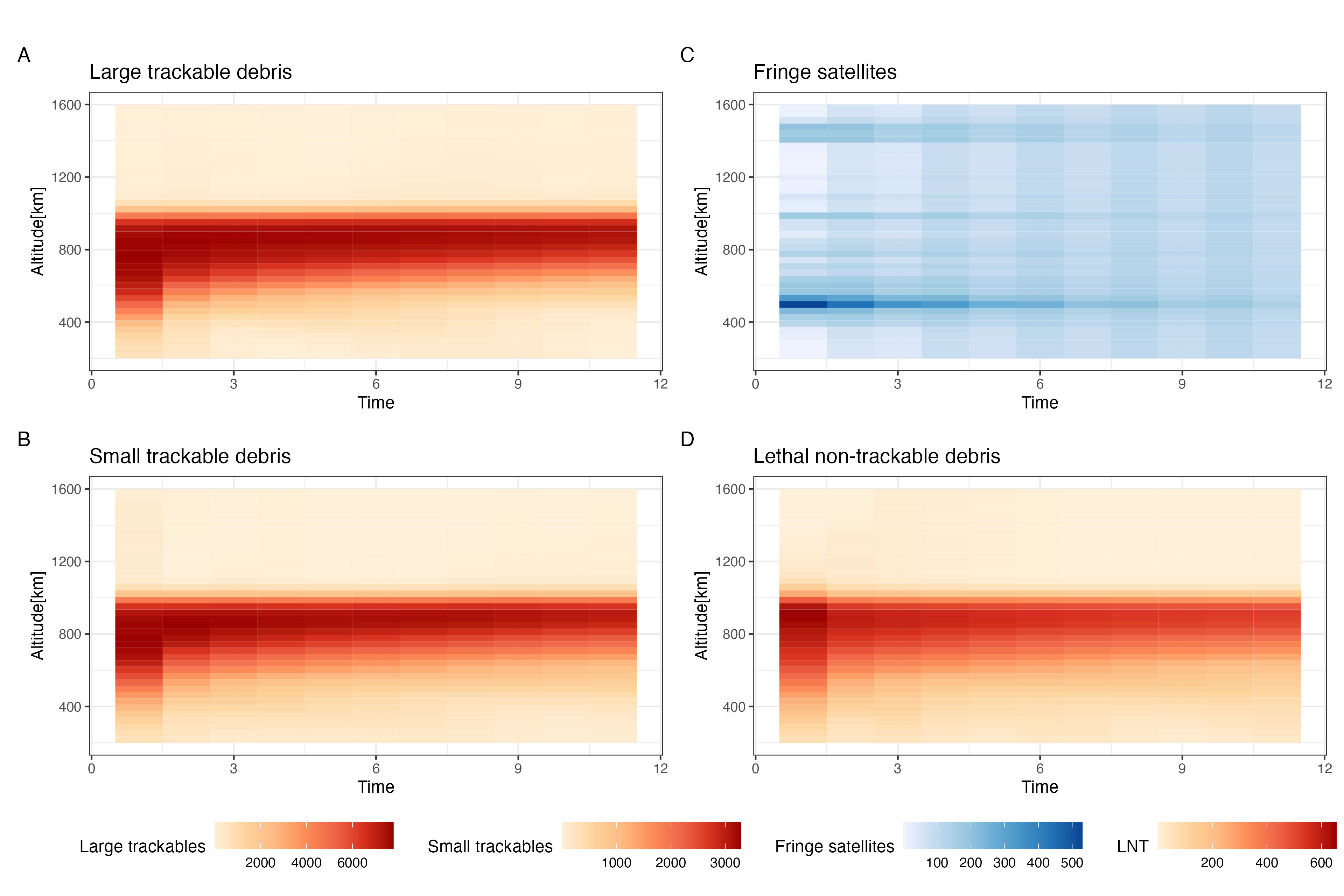}
    \caption{Total numbers of large trackable, small trackable, lethal non-trackable debris and fringe satellites from GMPHD filter propagation.}
    \label{fig:gmphd-test}
\end{figure}

Having demonstrated the potential of the OPUS framework to evaluate a diverse array of policies on equal footing, we demonstrate the framework's agnosticism to the specific propagator used. The GMPHD filter is still in early stages of development and requires further testing and validation before it can be used at the level of MOCAT-4S. However, we believe it shows promise, particularly for evaluating a larger state space of debris objects. To illustrate this potential we apply the GMPHD filter to evaluate the evolution of lethal non-trackable objects under equilibrium behavior. Since the GMPHD filter parameters have not been thoroughly tested, we run the simulation for only 10 years. Figure \ref{fig:gmphd-test} shows these results.

While the debris propagation appears to function effectively, the interaction with the economic solver produces unexpected results. After an initial period of plausible launch rates, the open-access condition appears to either have no interior solutions or become sufficiently irregular that the optimizer cannot find a solution, resulting in spatially-uniform oscillations. Regardless, the model is able to calculate implied counts of non-trackable debris in addition to larger trackable objects from propagating the underlying size distributions. These outcomes point to the need for further investigation of the GMPHD filter as an additional propagator to incorporate into the OPUS framework.

\section{Future research and development directions}
\label{sec:future}

The results presented above are a sample of what is possible in an IAM framework like OPUS. Using systems of equations that reflect payoffs from orbit use under different states of the world to predict launch patterns presents opportunities to study a wide array of policies, from disposal rules (e.g. 25- vs 5- year) to natural capital pricing (e.g. OUFs and performance bonds) to directed technology support (e.g. targeted debris removal subsidies). The framework also enables studying the effects of economic changes in the space sector, such as increased availability of smallsat launchers or new heavy-lift rockets. Below we outline a few directions for research and development to improve the utility of this IAM and others which may be developed in the future.

\paragraph{Improved economic parameter calibration.} A key limitation of current physico-economic modeling of orbit use is the lack of detailed economic data on the costs and revenues different operators incur/receive from using different locations. Such data is necessary to provide meaningful quantitative policy analyses. Projects like the Bureau of Economic Analysis' Space Economy satellite accounts are a useful step forward in this direction \citep{highfill2022estimating}, though they do not yet provide the level of detail on orbit use necessary for reliable quantitative estimates. Such data collection can also enable development of IAMs which incorporate linkages between orbit use and terrestrial economies, e.g. \cite{nozawa2023extent}.

\paragraph{Models of launch capacity and prices.} OPUS currently assumes open-access launchers are not constrained by limited rocket availability. This may be true at some points in time and not at others, though disentangling the causes of limited capacity---whether due to insufficient willingness-to-pay or fundamental constraints---is a separate question. Similarly, the price of launch is assumed to be constant throughout the simulation, though for policy analysis it may be useful to consider specific price trajectories, e.g. as in \cite{adilov2022analysis}. Constructing and incorporating models of the launch market would enable more realistic estimates of behavioral responses to policy proposals. Such models would also enable analysis of policies aimed at launch capacity, e.g. targeted support for particular types of launch vehicles.

\paragraph{More detailed models of operator behavior.} The economic behavior model applied here follows from the equilibrium conditions for launchers derived in \cite{rouillon2020physico, rao2020orbital, rao2023economics}. It is analytically convenient in studying launch intensity and location choices, as well as the overall impact of natural capital pricing. However, the present formulation only allows for a single per-satellite tax (``orbital-use fee''). It does not allow for easy differentiation between different implementations of natural capital pricing policies such as explicit satellite taxes \citep{rao2020orbital}, satellite deorbit performance bonds \citep{adilov2023economics}, or combinations of instruments intended to alter satellite design choices \citep{grzelka2019managing, guyot2023sustainable}. While these instruments can be described as \emph{types} of orbital-use fee, their implementations differ and can matter for overall policy effectiveness as well as implications for debris mitigation/remediation \citep{macauley2015economics, rao2018economic, guyot2023sustainable}.

\paragraph{Models of constellation behavior.} The model of launch behavior presented here focuses solely on operators who individually account for small shares of orbit use. Large constellations face different economic considerations and require different models. Recently there has been some progress in building economic models of constellations \citep{bernhard2023large, guyot2023satellite}, though these models do not incorporate open-access launchers. Incorporating models of constellation behavior into IAMs will enable more realistic and diverse policy analyses.

\paragraph{Models of non-commercial demand for satellite services.} Though commercial orbit use is growing, Figure \ref{fig:historical-patterns} shows that non-commercial orbit use remains a significant share of satellites in orbit. Many economically and socially important uses of orbital space are provided by government actors without commercial motives, e.g. GPS, and governments may function as an important source of demand for commercial services like telecommunications or remote sensing. The model of launch behavior presented here does not account for such patterns. However, the overall framework of OPUS can be extended to incorporate non-commercial orbit use. Such models may be easier to estimate from empirical data rather than derived from first principles, e.g. as in \cite{raoletizia2021}.

\section{Conclusion}
\label{sec:conclusion}

In this paper we have introduced and validated an Integrated Assessment Model (IAM) capable of analyzing both the physical behavior of objects in orbit and the economic behavior of entities controlling these objects. By combining orbital propagators with economic modeling, OPUS provides a robust tool that can help policymakers better understand and mitigate the challenges associated with the growing congestion of Earth's orbital space.

Many types of analyses are possible within the OPUS framework. These range from studying the environmental effects of reductions in the cost of launching or operating a satellite to exploring the implications of increases in commercial demand for satellite services. OPUS can also be applied to investigate the effects of anti-satellite (ASAT) tests, among other policy-relevant scenarios. While OPUS' capabilities have been demonstrated, it is worth emphasizing that the results presented here are intended to illustrate the model's versatility and potential applications rather than provide specific policy guidance.

There are several limitations in the current implementation of OPUS. A major limitation is the lack of detailed economic data on the costs and revenues for different satellite operators. Furthermore, OPUS currently only endogenizes open access to orbit by commercial launchers, with large constellation behavior treated as exogenous and non-commercial use not included. Future research could extend OPUS to include models focusing on large constellations and non-commercial demands for satellite services. Despite its limitations, OPUS is a useful step toward a holistic understanding of space sustainability issues. Policymakers can use this tool to weigh the costs and benefits of various proposals, from orbital slotting concepts and deorbit timelines to targeted debris removal subsidies and orbital-use fees.

\newpage

{
	\setlength{\bibsep}{3pt}
	\setstretch{1}
	\bibliography{bibliography}
	\bibliographystyle{jpe}
}

\end{document}